\documentclass[aps,prd,twocolumn,amsmath,amssymb,amsfonts,nofootinbib,superscriptaddress,altaffilletter]{revtex4-1}

\usepackage{graphicx}
\usepackage{hyperref}
\hypersetup{
  bookmarksopen=true
}
\usepackage{amssymb}
\usepackage{amsmath}
\usepackage{color}
\usepackage{supertabular}
\usepackage{dcolumn}
\usepackage{multirow}

\def\beq{\begin{equation}}
\def\eeq{\end{equation}}
\def\bea{\begin{eqnarray}}
\def\eea{\end{eqnarray}}

\def\max{\textrm{\mbox{\tiny{max}}}}

\def\ec{\textrm{~,}}
\def\ed{\textrm{~.}}
\def\fdot{f^{(1)}}

\newif\ifshowfigs
\showfigstrue

\newcommand{\SNR}{\textrm{SNR}}
\newcommand{\Fstat}{$\mathcal{F}$-statistic}
\newcommand{\nseg}{$N_{\textrm{seg}}$}
\newcommand\T{\rule{0pt}{2.6ex}}       
\newcommand\B{\rule[-1.2ex]{0pt}{0pt}} 

\def\pf{{\footnotesize\textsc{PowerFlux}}}
\def\pyfstat{{\footnotesize\textsc{PyFstat}}}

\def\fgw{f_{\rm GW}}
\def\frot{f_{\rm rot}}
\def\Izz{I_{\rm zz}}

\def\sci#1#2{#1\times10^{#2}}

\def\etal{{\it et al.}}
\def\vs{{\it vs.}}

\def\RAJ{\textrm{RA}_{\textrm J2000}}
\def\DECJ{\textrm{DEC}_{\textrm J2000}}

\newcounter{BandAcountzero} \newcounter{BandBcountzero}\newcounter{BandCcountzero} \newcounter{BandDcountzero}\newcounter{BandEcountzero}
\newcounter{BandFcountzero} \newcounter{BandGcountzero}\newcounter{BandHcountzero} \newcounter{BandIcountzero}\newcounter{BandJcountzero}
\newcounter{BandKcountzero} \newcounter{BandLcountzero}\newcounter{BandMcountzero} \newcounter{BandNcountzero}\newcounter{BandOcountzero}
\newcounter{BandPcountzero} \newcounter{BandQcountzero}\newcounter{BandRcountzero} \newcounter{BandScountzero}\newcounter{BandTcountzero}
\newcounter{VlowBandcountzero} \newcounter{LowBandcountzero} \newcounter{MediumBandcountzero} \newcounter{HighBandcountzero}

\newcounter{BandAcountone} \newcounter{BandBcountone}\newcounter{BandCcountone} \newcounter{BandDcountone}\newcounter{BandEcountone}
\newcounter{BandFcountone} \newcounter{BandGcountone}\newcounter{BandHcountone} \newcounter{BandIcountone}\newcounter{BandJcountone}
\newcounter{BandKcountone} \newcounter{BandLcountone}\newcounter{BandMcountone} \newcounter{BandNcountone}\newcounter{BandOcountone}
\newcounter{BandPcountone} \newcounter{BandQcountone}\newcounter{BandRcountone} \newcounter{BandScountone}\newcounter{BandTcountone}
\newcounter{VlowBandcountone} \newcounter{LowBandcountone} \newcounter{MediumBandcountone} \newcounter{HighBandcountone}

\newcounter{BandAcounttwo} \newcounter{BandBcounttwo}\newcounter{BandCcounttwo} \newcounter{BandDcounttwo}\newcounter{BandEcounttwo}
\newcounter{BandFcounttwo} \newcounter{BandGcounttwo}\newcounter{BandHcounttwo} \newcounter{BandIcounttwo}\newcounter{BandJcounttwo}
\newcounter{BandKcounttwo} \newcounter{BandLcounttwo}\newcounter{BandMcounttwo} \newcounter{BandNcounttwo}\newcounter{BandOcounttwo}
\newcounter{BandPcounttwo} \newcounter{BandQcounttwo}\newcounter{BandRcounttwo} \newcounter{BandScounttwo}\newcounter{BandTcounttwo}
\newcounter{VlowBandcounttwo} \newcounter{LowBandcounttwo} \newcounter{MediumBandcounttwo} \newcounter{HighBandcounttwo}

\newcounter{VlowBandcountHWinj} \newcounter{LowBandcountHWinj} \newcounter{MediumBandcountHWinj} \newcounter{HighBandcountHWinj}
\newcounter{VlowBandcountStrainhist} \newcounter{LowBandcountStrainhist} \newcounter{MediumBandcountStrainhist} \newcounter{HighBandcountStrainhist}
\newcounter{VlowBandcountFullO3} \newcounter{LowBandcountFullO3} \newcounter{MediumBandcountFullO3} \newcounter{HighBandcountFullO3}

\addtocounter{LowBandcountzero}{\value{BandCcountzero}} \addtocounter{LowBandcountzero}{\value{BandDcountzero}} 
\addtocounter{LowBandcountzero}{\value{BandEcountzero}}

\addtocounter{MediumBandcountzero}{\value{BandGcountzero}} \addtocounter{MediumBandcountzero}{\value{BandHcountzero}}
\addtocounter{MediumBandcountzero}{\value{BandIcountzero}} \addtocounter{MediumBandcountzero}{\value{BandJcountzero}}
\addtocounter{MediumBandcountzero}{\value{BandKcountzero}} \addtocounter{MediumBandcountzero}{\value{BandLcountzero}}
\addtocounter{MediumBandcountzero}{\value{BandMcountzero}} \addtocounter{MediumBandcountzero}{\value{BandNcountzero}}
\addtocounter{MediumBandcountzero}{\value{BandOcountzero}}

\addtocounter{HighBandcountzero}{\value{BandQcountzero}} \addtocounter{HighBandcountzero}{\value{BandRcountzero}}
\addtocounter{HighBandcountzero}{\value{BandScountzero}} \addtocounter{HighBandcountzero}{\value{BandTcountzero}}

\addtocounter{LowBandcountone}{\value{BandCcountone}} \addtocounter{LowBandcountone}{\value{BandDcountone}}
\addtocounter{LowBandcountone}{\value{BandEcountone}}

\addtocounter{MediumBandcountone}{\value{BandGcountone}} \addtocounter{MediumBandcountone}{\value{BandHcountone}}
\addtocounter{MediumBandcountone}{\value{BandIcountone}} \addtocounter{MediumBandcountone}{\value{BandJcountone}}
\addtocounter{MediumBandcountone}{\value{BandKcountone}} \addtocounter{MediumBandcountone}{\value{BandLcountone}}
\addtocounter{MediumBandcountone}{\value{BandMcountone}} \addtocounter{MediumBandcountone}{\value{BandNcountone}}
\addtocounter{MediumBandcountone}{\value{BandOcountone}}

\addtocounter{HighBandcountone}{\value{BandQcountone}} \addtocounter{HighBandcountone}{\value{BandRcountone}}
\addtocounter{HighBandcountone}{\value{BandScountone}} \addtocounter{HighBandcountone}{\value{BandTcountone}}

\addtocounter{LowBandcounttwo}{\value{BandCcounttwo}} \addtocounter{LowBandcounttwo}{\value{BandDcounttwo}}
\addtocounter{LowBandcounttwo}{\value{BandEcounttwo}}

\addtocounter{MediumBandcounttwo}{\value{BandGcounttwo}} \addtocounter{MediumBandcounttwo}{\value{BandHcounttwo}}
\addtocounter{MediumBandcounttwo}{\value{BandIcounttwo}} \addtocounter{MediumBandcounttwo}{\value{BandJcounttwo}}
\addtocounter{MediumBandcounttwo}{\value{BandKcounttwo}} \addtocounter{MediumBandcounttwo}{\value{BandLcounttwo}}
\addtocounter{MediumBandcounttwo}{\value{BandMcounttwo}} \addtocounter{MediumBandcounttwo}{\value{BandNcounttwo}}
\addtocounter{MediumBandcounttwo}{\value{BandOcounttwo}}

\addtocounter{HighBandcounttwo}{\value{BandQcounttwo}} \addtocounter{HighBandcounttwo}{\value{BandRcounttwo}}
\addtocounter{HighBandcounttwo}{\value{BandScounttwo}} \addtocounter{HighBandcounttwo}{\value{BandTcounttwo}}

\def\VlowBandcountzero{{\color{magenta}\arabic{VlowBandcountzero}}}
\def\LowBandcountzero{{\color{magenta}\arabic{LowBandcountzero}}}
\def\MediumBandcountzero{{\color{magenta}\arabic{MediumBandcountzero}}}
\def\HighBandcountzero{{\color{magenta}\arabic{HighBandcountzero}}}

\def\VlowBandcountone{{\color{magenta}\arabic{VlowBandcountone}}}
\def\LowBandcountone{{\color{magenta}\arabic{LowBandcountone}}}
\def\MediumBandcountone{{\color{magenta}\arabic{MediumBandcountone}}}
\def\HighBandcountone{{\color{magenta}\arabic{HighBandcountone}}}

\def\VlowBandcounttwo{{\color{magenta}\arabic{VlowBandcounttwo}}}
\def\LowBandcounttwo{{\color{magenta}\arabic{LowBandcounttwo}}}
\def\MediumBandcounttwo{{\color{magenta}\arabic{MediumBandcounttwo}}}
\def\HighBandcounttwo{{\color{magenta}\arabic{HighBandcounttwo}}}

\def\VlowBandcountHWinj{{\color{magenta}\arabic{VlowBandcountHWinj}}}
\def\LowBandcountHWinj{{\color{magenta}\arabic{LowBandcountHWinj}}}
\def\MediumBandcountHWinj{{\color{magenta}\arabic{MediumBandcountHWinj}}}
\def\HighBandcountHWinj{{\color{magenta}\arabic{HighBandcountHWinj}}}

\def\VlowBandcountStrainhist{{\color{magenta}\arabic{VlowBandcountStrainhist}}}
\def\LowBandcountStrainhist{{\color{magenta}\arabic{LowBandcountStrainhist}}}
\def\MediumBandcountStrainhist{{\color{magenta}\arabic{MediumBandcountStrainhist}}}
\def\HighBandcountStrainhist{{\color{magenta}\arabic{HighBandcountStrainhist}}}

\def\VlowBandcountFullO3{{\color{magenta}\arabic{VlowBandcountFullO3}}}
\def\LowBandcountFullO3{{\color{magenta}\arabic{LowBandcountFullO3}}}
\def\MediumBandcountFullO3{{\color{magenta}\arabic{MediumBandcountFullO3}}}
\def\HighBandcountFullO3{{\color{magenta}\arabic{HighBandcountFullO3}}}

\begin{document}

\title{
All-sky Search for Continuous Gravitational Waves from Isolated Neutron Stars in the Early O3 LIGO Data
}

\let\mymaketitle\maketitle
\let\myauthor\author
\let\myaffiliation\affiliation
\author{LIGO Scientific Collaboration}
\author{Virgo Collaboration}
\author{KAGRA Collaboration}

\date[\relax]{compiled \today}

\noaffiliation

\begin{abstract}
  We report on an all-sky search for continuous gravitational waves in the frequency band 20-2000\,Hz
  and with a frequency time derivative in the range of $\sci{[-1.0, +0.1]}{-8}$\,Hz/s. Such a signal could be produced by
  a nearby, spinning and slightly non-axisymmetric isolated neutron star in our galaxy.
  This search uses the LIGO data from the first six months of Advanced LIGO's and Advanced Virgo's third observational run, O3.
  No periodic gravitational wave signals are observed, and 95\%\ confidence-level (CL) frequentist upper limits are placed on their strengths.
  The lowest upper limits on worst-case (linearly polarized) strain amplitude $h_0$ are  $\sim\sci{1.7}{-25}$ near 200\,Hz.
  For a circularly polarized source (most favorable orientation), the lowest
  upper limits are  $\sim\sci{6.3}{-26}$.
  These strict frequentist upper limits refer to all sky locations and the entire range of frequency
  derivative values.
  For a population-averaged ensemble of sky locations and stellar orientations,
  the lowest 95\%\ CL upper limits on the strain amplitude are $\sim\sci{1.4}{-25}$.
  These upper limits improve upon our previously published all-sky results,
  with the greatest improvement (factor of $\sim$2) seen at higher frequencies, in part because quantum squeezing
  has dramatically improved the detector noise level relative to the second observational run, O2.
  These limits are the most constraining to date over most of the parameter space searched.
    
\end{abstract}

\maketitle

\section{Introduction}
\label{sec:introduction}

We report the results of an all-sky search for continuous, nearly monochromatic gravitational waves from
rapidly rotating isolated neutron stars using the first six months of data from the third observing run (O3) of the
Advanced Laser Interferometer Gravitational wave Observatory (Advanced LIGO \cite{bib:LIGO,bib:DetectorPaper}). This first search of the
early O3 data uses the \pf\ hierarchical search program~\cite{bib:S4allsky,bib:S5allsky1,bib:S5allsky2,bib:S6allsky,bib:O1allsky1,bib:O1allsky2}
with loose-coherence follow-up~\cite{bib:loosecoherence1,bib:loosecoherence2} of outliers and covers a broad band of frequencies
(20-2000 Hz) and frequency derivatives ($\sci{[-1.0, +0.1]}{-8}$~Hz/s).
Although more than $\sci{1.4}{5}$ search outliers are found in the initial stage of analysis, successive follow-up stages failed to
confirm an astrophysical signal. Hence the primary results from this analysis are the
upper limits on strain amplitude presented in section~\ref{sec:upperlimits}.

All-sky searches for continuous gravitational waves from isolated neutron stars
have been carried out in Advanced LIGO and Virgo data
previously~\cite{bib:O1allsky1,bib:O1allsky2,bib:O1EatH,bib:O2allsky,bib:O2allskyRome,bib:O2EatH,bib:O2Falcon1,bib:O2Falcon2,bib:O2Falcon3,bib:deepsearch171-172Hz}.
The results presented here are the most sensitive to date in strain amplitude for a broadband, all-sky search with high allowed
spin-down magnitudes, with improvement factors ranging from $\sim$1.2 at signal frequencies of
$\sim$100 Hz to $\sim$2 at the higher frequencies. 
The improvement with respect to similar, previous analyses of O2 data~\cite{bib:O2allsky,bib:O2allskyRome}
is greater at higher frequencies, in part because of the improved detector noise ($\sim$3db) achieved with
quantum squeezing~\cite{bib:O3squeezing} and in part because 
longer Fourier transform coherence times are used here (for frequencies up to 1475 Hz) than in the O2 analyses.

Our primary targets in this analysis are fast-spinning, non-axisymmetric neutron stars in the Milky Way.
Given the immense pressure on its nuclear matter, one expects a neutron star to assume a
highly spherical shape in the limit of no rotation and, with rotation, to form
an axisymmetric oblate spheroid. A number of physical processes can disrupt the symmetry,
however, to produce quadrupolar gravitational waves from the stellar rotation.
Those processes include crustal distortions from cooling or accretion,
buried magnetic field energy and excitation of r-modes. Comprehensive reviews of
continuous gravitational wave emission mechanisms from neutron stars can be found in \cite{bib:Lasky_review,bib:GandG_review}

This article is organized as follows: Section~\ref{sec:data} describes the data set used, including steps taken to
mitigate extremely loud and relatively frequent instrumental glitches seen in the O3 LIGO data,
a phenomenon not seen in previous LIGO observing runs. Section~\ref{sec:method} briefly describes the
\pf\ and loose-coherence algorithm used in this and previous searches. Section~\ref{sec:results} presents
the results of the analysis. Section~\ref{sec:conclusions} concludes with a discussion of the results and
prospects for future searches.

\section{Data sets used}
\label{sec:data}

Advanced LIGO consists of two detectors, one in Hanford, Washington (designated H1), and the other in Livingston, Louisiana (designated L1), separated by a $\sim$3000-km baseline~\cite{bib:LIGO}.
Each site hosts one, 4-km-long interferometer inside a vacuum envelope with the primary interferometer optics suspended by a cascaded, quadruple suspension system in order to isolate them from external disturbances.
The interferometer mirrors act as test masses, and the passage of a gravitational wave induces a differential-arm length change that is proportional to the gravitational-wave strain amplitude.

Advanced LIGO's first observing run (O1) occurred between September 12, 2015 and
January 19, 2016 and led to the discovery on September 14, 2015 of gravitational waves
from binary black hole (BBH) coalescences~\cite{bib:GW150914}.
The O2 observing run began November 30, 2016 and ended August 25, 2017 and included the first detection
of a binary neutron star (BNS) merger~\cite{bib:GW170817}.
The O3 run began April 1, 2019 and ended March 27, 2020, for which the first six months (April 1, 2019 to October 1, 2019),
prior to a 1-month commissioning break, is designated as the O3a epoch.
The analysis presented here is based primarily on the O3a data set, with data from the remainder of the run (O3b epoch)
used only for following up on promising signal candidates.
From the O1, O2 and O3a data sets, the LIGO Scientific Collaboration
and Virgo Collaboration have reported a total of 50 compact binary coalescences~\cite{bib:O1catalog,bib:O2catalog,bib:O3acatalog},
primarily BBH events, with two binary neutron star (BNS) detections~\cite{bib:GW170817,bib:GW190425}
and one potential binary neutron star / black hole detection~\cite{bib:GW190814}.

The Virgo interferometer~\cite{bib:Virgo} observed during August 2017 near the end of the O2 run
and throughout the O3 run. The Virgo data has not
been used in this analysis, however, because of an unattractive tradeoff in computational cost for sensitivity gain, given
the interferometer's higher noise level during the O3 run.

Prior to searching the O3a data for continuous wave (CW) signals, the quality of the data was assessed
and steps taken to mitigate the effects of instrumental artifacts. As in previous Advanced LIGO observing
runs~\cite{bib:linespaper}, instrumental ``lines'' (sharp peaks in fine-resolution run-averaged H1 and L1 spectra)
are marked, and where
possible, their instrumental or environmental sources identified~\cite{bib:O3aLinelist}. The resulting database of artifacts proved helpful in
eliminating spurious signal candidates emerging from the search. In general, the line multiplicity for H1 O3a data was
similar to that observed in the O2 run, while the line multiplicity for L1 O3a data was substantially reduced.

Another type of artifact observed in the O3a data for both H1 and L1 were relatively frequent and loud
``glitches'' (short, high-amplitude instrumental transients) with most of their spectral power lying below $\sim$500 Hz.
Although loud glitches have been observed in previous runs, their frequency in O3a was dramatically higher.
At present, investigations of the source of these glitches remain inconclusive. An effort to identify and mitigate
them during the October 2019 commissioning break (between the O3a and O3b epochs) did not succeed.
Unlike in previous LIGO data runs, the sheer spectral power in the glitches increased the effective broadband noise
floor of the data below $\sim$500 Hz quite substantially, as seen in run-averaged spectra computed from 1800s and 7200s
discrete Fourier transforms known as ``SFTs'' (for ``short Fourier transforms''). Most CW searches based on
summing strain spectral power from SFTs, including the \pf\ program used here,
apply weightings that disfavor SFTs with high noise. Because the average time interval between loud glitches
in O3a data is comparable to or smaller than the coherence time of the SFTs, however, inverse noise weighting of SFTs
proved much less effective than in previous runs, especially for the 7200s SFTs.

To mitigate the effects of these glitches on O3a CW searches for signals below 475 Hz, a simple glitch-gating
algorithm was applied ~\cite{bib:gatingdoc,bib:detcharpaper} to excise the transients from the data. 
For each 1/16-second for which a whitened version of the H1 and L1 strain data channels
had excess RMS power in the 25-50 Hz or 70-110 Hz bands, the strain channel was set to zero. To reduce artifacts
from discontinuous data, a 1/4-second half-Hann-window ramping from unity to zero was multiplied against the data stream
prior to each zeroed interval, and a 1/4-second half-Hann window ramping from zero back to unity was multiplied at the end
of zeroed intervals. This gating can be considered to be inverse-Tukey-windowed. Although the Tukey windowing
mitigates severe spectral artifacts, the resulting spectra still suffer visible spectral leakage
very near loud instrumental lines, such as from 60-Hz power mains and ``violin modes'' (near 500 Hz) due to
ambient vibrations of the silica fibers from which LIGO mirrors are suspended.
All gated intervals longer than 30 seconds are excluded from analysis, as are 7200s SFTs 
containing total gate durations longer than 120 seconds. Altogether, the applied gating leads in this analysis to losses
of about 1\%\ and 11\%\ of the H1 and L1 observation times during the O3a epoch for search frequencies below 475 Hz.
Details of the gating, including validation that
low-frequency CW ``hardware injections'' (simulated signals imposed on interferometer mirrors during data taking, see
Section~\ref{sec:hwinjections}) are recovered with higher signal-to-noise ratio (SNR) in gated data than in the
original, ungated data can be found in a technical report~\cite{bib:gatingdoc}.

As discussed in section~\ref{sec:method}, the \pf\ program searches many narrow frequency bands, where the SFT coherence time chosen depends on frequency. In general, longer coherence times give
improved sensitivity, but incur larger computational costs from the need to search more finely in parameter space.
In addition, efficiency loss from spectral leakage due to Doppler modulations from the Earth's motion increases
at higher frequency, especially for sources near the ecliptic plane. Given these tradeoffs, the same coherence-time
choices made for the O1 analysis~\cite{bib:O1allsky2} are chosen here, as shown in Table~\ref{tab:sft_info}.
The SFTs are created from the C01 calibrated strain data~\cite{bib:O3calibpaper}, using Hann windowing and 50\%\ overlap.
Upper limits on the calibration uncertainties over the 20-2000 Hz band in the O3a epoch are estimated
to be $<$ 7\%\ in magnitude and $<$ 4 deg in phase (68\% confidence interval).

\begin{table}[htb]
\begin{center}
  \begin{tabular}{lccc}\hline
    \T\B                                &  20-475 Hz      & 475-1475 Hz       & 1475-2000 Hz \\
    \hline\hline
    \T\B Data set\strut                       & Gated C01       & Ungated C01       & Ungated C01  \\
    \hline
    \T\B Coherence time\strut             & 7200s           & 3600s             & 1800s        \\
    \hline
    \multirow{2}*{Number of SFTs}  & H1:\T 2287        & H1: 5535          & H1: 11735     \\
                                   & L1:\B 2247        & L1: 5973          & L1: 12574     \\
\hline 
\end{tabular}
  \caption{Information on the Short Fourier Transforms (SFTs) used in the search. All SFTs are
    Hann-windowed with 50\%\ overlap. The numbers of gated SFTs used below 475 Hz are affected
    by livetime loss from avoiding data stretches with high rates of gated gltiches, with a
    larger effect seen for L1 data.}
\label{tab:sft_info}
\end{center}
\end{table}

\section{Analysis Method}
\label{sec:method}

This search uses the \pf\ program~\cite{bib:S4allsky,bib:O1allsky1,bib:O1allsky2,bib:S5allsky2} with loose-coherence follow-up
of outliers~\cite{bib:loosecoherence1,bib:loosecoherence2}. In brief, strain power is summed over many SFTs after
correcting for Doppler modulations, for a large bank of templates based on sky location, frequency, frequency
derivative and stellar orientation. The maximum strain powers detected over the entire sky and for all
frequency derivatives in each narrow frequency sub-band (see Table~\ref{tab:tuning_info}
for the frequency-dependent widths) define strict frequentist upper limits.
Initial outliers, defined by high values of SNR, are followed up in a
multi-stage search with increasing levels of effective coherence at each stage
to increase the expected SNR for true signals. Candidates passing all stages are
examined individually to determine their credibility as gravitational wave sources,
with further warranted follow-up carried out in the full O3 data set.

\subsection{Signal model and parameter space searched}
\label{sec:model}

The signal templates assume a classical model of a spinning neutron star with a time-varying quadrupole moment that produces
circularly polarized gravitational radiation along the rotation axis, linearly polarized radiation
in the directions perpendicular to the rotation axis and elliptical polarization for the general case.
The star's orientation, which determines the polarization, is parametrized by the inclination angle $\iota$ of its
spin axis relative to the detector line-of-sight and by the angle $\psi$ of the axis projection on the plane of the sky.
The linear polarization case ($\iota=\pi/2$) is the most unfavorable 
because the gravitational wave flux impinging on the detectors
is smallest for an intrinsic strain amplitude $h_0$,
possessing eight times less incident strain power than
for circularly polarized waves ($\iota = 0,\>\pi$).

The strain signal model $h(t)$ for a periodic source is assumed to be the following function of time $t$:
\begin{eqnarray}
  h(t) & = & h_0\bigl(F_+(t, \alpha_0, \delta_0, \psi)\frac{1+\cos^2(\iota)}{2}\cos(\Phi(t))\vphantom{\frac{1+\cos^2(\iota)}{2}}\nonumber\\
  & & \quad\>+F_\times(t, \alpha_0, \delta_0, \psi)\cos(\iota)\sin(\Phi(t))\bigr)\ec
\end{eqnarray}

\noindent where $h_0$ is the intrinsic strain amplitude, $\Phi(t)$ is the signal phase, $F_+$ and $F_\times$ characterize the detector responses to signals with ``$+$'' and ``$\times$''
quadrupolar polarizations \cite{bib:S4allsky}, and the sky location is described by right ascension $\alpha_0$ and
declination $\delta_0$.

In a rotating triaxial ellipsoid model for a 
star at distance $r$ spinning at frequency $\frot$ about its (approximate) symmetry axis ($z$),
the amplitude $h_0$ can be expressed as
\begin{eqnarray}
  \label{eqn:hexpected}
h_0 & = & {4\,\pi^2G\epsilon\Izz\fgw^2\over c^4r}  \\
    & = & [1.1\times10^{-24}]\Bigl[{\epsilon\over10^{-6}}\Bigr]\Bigl[{\Izz\over I_0}\Bigr]\Bigl[{\fgw\over1\>{\rm kHz}}\Bigr]^2
\Bigl[{1\>{\rm kpc}\over r}\Bigr],
\end{eqnarray}
where $I_0=10^{38}$ kg$\cdot$m$^2$ (10$^{45}$ g$\cdot$cm$^2$) is a nominal neutron star moment of inertia $\Izz$ about $z$,
and the gravitational radiation is emitted at frequency $\fgw=2\,\frot$.
The equatorial ellipticity $\epsilon$ is a convenient, dimensionless measure of stellar non-axisymmetry:
\begin{equation}
\label{eqn:ellipticity}
\epsilon \quad \equiv \quad {I_{xx}-I_{yy}\over I_{zz}}.
\end{equation}

The phase evolution of the signal is given in the reference frame of the Solar System barycenter (SSB) by the second-order approximation:
\begin{equation}
\label{eqn:phase_evolution}
\Phi(t)=2\pi\left(f_\textrm{source}\cdot (t-t_0)+\fdot\cdot (t-t_0)^2/2\right)+\phi\ec
\end{equation}
where $f_\textrm{source}$ is the SSB source frequency, $\fdot$ is the first frequency derivative (which, when negative,
is termed the spin-down), $t$ is the SSB time, and the initial phase $\phi$ is
computed relative to reference time $t_0$. When expressed as a function of the local time of ground-based detectors,
Eq.~\ref{eqn:phase_evolution} acquires sky-position-dependent Doppler shift terms.

We search a frequency band 20-2000 Hz and a frequency derivative $\dot{f}$ range
of $[-\sci{1}{-8},+\sci{1}{-9}]~\mathrm{Hz}~\mathrm{s}^{-1}$.
Figure~\ref{fig:parameterspace} shows this parameter space coverage together with
those of previous all-sky searches of the LIGO O2 data using different methods. 
Most natural sources are expected to have a negative first frequency derivative, as the energy lost in gravitational or
electromagnetic waves would make the source spin more slowly.
A small number of isolated pulsars in globular clusters exhibit slight apparent spin-up, 
believed to arise from acceleration in the Earth's direction;  known apparent spin-up values have
magnitudes too small to prevent source detection with the zero-spin-down templates
used in this search, given a strong enough signal.
The frequency derivative can also be positive when the source
is affected by a strong slowly-varying Doppler shift, such as due to a long-period orbit with a companion.
A more exotic source of spin-up is gravitational wave superradiance from a boson cloud
in the vicinity of an isolated black hole~\cite{bib:bosoncloud}.

All known isolated pulsars spin down more slowly
than the maximum value of $|\dot{f}|_\max$ used here, and as seen in the results
section, the equatorial ellipticity required for higher $|\dot{f}|$ is improbably high
for a source losing rotational energy primarily via gravitational radiation at low
frequencies. More plausible is a source with spin-down dominated by electromagnetic
radiation energy loss, but for which detectable gravitational radiation is also emitted.

One measure of gravitational wave detectability of known pulsars is the ``spin-down strain limit"
defined by equating inferred rotational kinetic energy loss to gravitational wave energy emission.
For the frequency band searched here, such limits range from as high as O($10^{-24}$) for energetic,
low-frequency young stars, such as the Crab and Vela pulsars, down to below $10^{-27}$ for high-frequency
millisecond pulsars~\cite{bib:O2targetedpaper}, where the highest such limit at high frequencies
is O($10^{-26}$). 

\subsection{Methodology}
\label{sec:methodology}

The \pf\ pipeline has a hierarchical structure
that permits systematic follow-up of loud outliers from the
initial stage. The later stages improve intrinsic strain sensitivity by
increasing effective coherence time while dramatically
reducing the parameter space volume over which the follow-up
is pursued. The pipeline uses loose coherence~\cite{bib:loosecoherence1}
with stages of improving refinement via steadily increasing effective
coherence times. Any outliers that survive
all stages of the search pipeline are examined manually for
contamination from known instrumental artifacts and for evidence of
contamination from a previously unknown single-interferometer artifact.
Those for which no artifacts are found are subjected to further
follow-up described below.

In the pipeline's initial stage, the main \pf\
algorithm \cite{bib:S4allsky,bib:O1allsky1,bib:O1allsky2,bib:S5allsky2,bib:S5allsky1,bib:S6allsky}
establishes upper limits and produces lists of outliers.
The program sets strict frequentist upper limits
on detected strain power in circular and linear polarizations that
apply everywhere on the sky except for small regions near the ecliptic
poles, where signals with small Doppler modulations can be masked by
stationary instrumental spectral lines.
The procedure defining these excluded regions is described in \cite{bib:S5allsky2} and applies to less than $0.2$\% of the sky over the entire run,
where the precise shapes of the regions near the poles depend on assumed signal frequency and spin-down.
Initial outliers are defined by a 
joint H1-L1 signal-to-noise ratio (SNR) greater than a threshold of 7, with consistency among
corresponding H1, L1 and joint H1-L1
outliers (criteria described in section~\ref{sec:outlier_followup}). These outliers are then followed up with
a loose-coherence detection pipeline \cite{bib:loosecoherence1,bib:loosecoherence2,bib:S5allsky2}, which is
used to reject or confirm the outliers.

The power calculation of the data can be expressed as a bilinear form of the input matrix $\left\{a_{t,f}\right\}$
constructed from the SFT coefficients with indices representing time and frequency:

\begin{equation}
P[f] = \sum_{t_1, t_2,D_i,D_j} a_{t_1, f+\Delta f(t_1)}^{(D_i)} a_{t_2, f+\Delta f(t_2)}^{(D_j)*} K_{t_1, t_2, f}^{D_iD_j}\ed
\end{equation}

\noindent In this expression $\Delta f(t)$ is the detector-frame frequency drift due to the effects from both Doppler shifts and the first
frequency derivative. The sum is taken over all times $t$ corresponding to the midpoints of the SFT time intervals.
The kernel $K_{t_1, t_2, f}^{(D_iD_j)}$ includes the contribution of time-dependent SFT noise weights, antenna response, signal
polarization parameters, and relative phase terms~\cite{bib:loosecoherence1,bib:loosecoherence2} for detectors $D_{i,j}$ (= H1, L1).
Separate power sums are computed for H1, L1 and combined H1-L1 data.

The fast first-stage (stage 0) \pf\ algorithm uses a kernel with diagonal terms only (including separate single-detector
contributions $D_i=D_j$). The second stage (stage 1)
increases effective coherence time while still allowing for controlled deviation in
phase~\cite{bib:loosecoherence1} via kernels that increase effective coherence length by inclusion of
limited single-detector, off-diagonal terms. The third stage (stage 2) maintains the stage-1 effective coherence time, but
adds SFT coefficients from H1 and L1 data coherently ($D_i\ne D_j$) to improve SNR and parameter resolution.

The effective coherence length is captured in a parameter $\delta$~\cite{bib:loosecoherence1}, which describes the degree of phase drift 
allowed between SFTs. A value of $\delta=0$ corresponds to a fully coherent case,
and $\delta=\pi$ corresponds to incoherent power sums. 

Depending on the terms used, the data from different interferometers can be combined incoherently (such as in stages 0
and 1, see Table \ref{tab:followup_parameters}) or coherently (as used in stage 2). The coherent
combination is more computationally expensive but improves parameter estimation. 

\subsection{Upper limits determination}
\label{sec:upperlimit_determination}
  
The 95\% confidence-level (CL) upper limits presented in section~\ref{sec:results}
are reported in terms of the worst-case value of $h_0$ (linear polarization) and for
the most sensitive case of circular polarization.

These upper limits,
produced in stage 0, are based
on the overall noise level and largest outlier in strain found for every template in each narrow sub-band in
the first stage of the pipeline. Sub-bands are analyzed by separate instances of \pf~\cite{bib:S5allsky2}

\begin{figure*}[htbp]
  \begin{center}
    \ifshowfigs
    \includegraphics[width=3.3in]{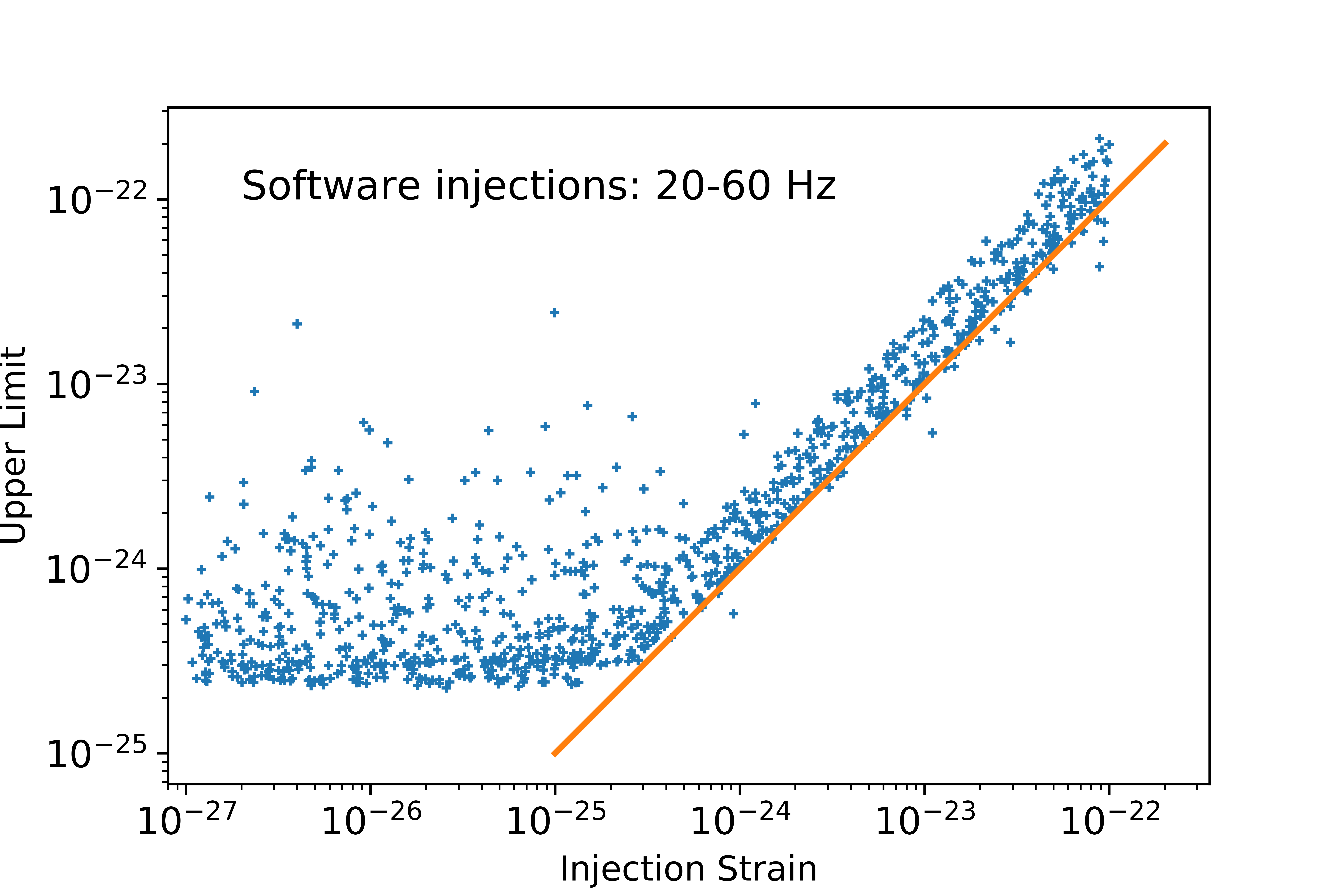}
    \includegraphics[width=3.3in]{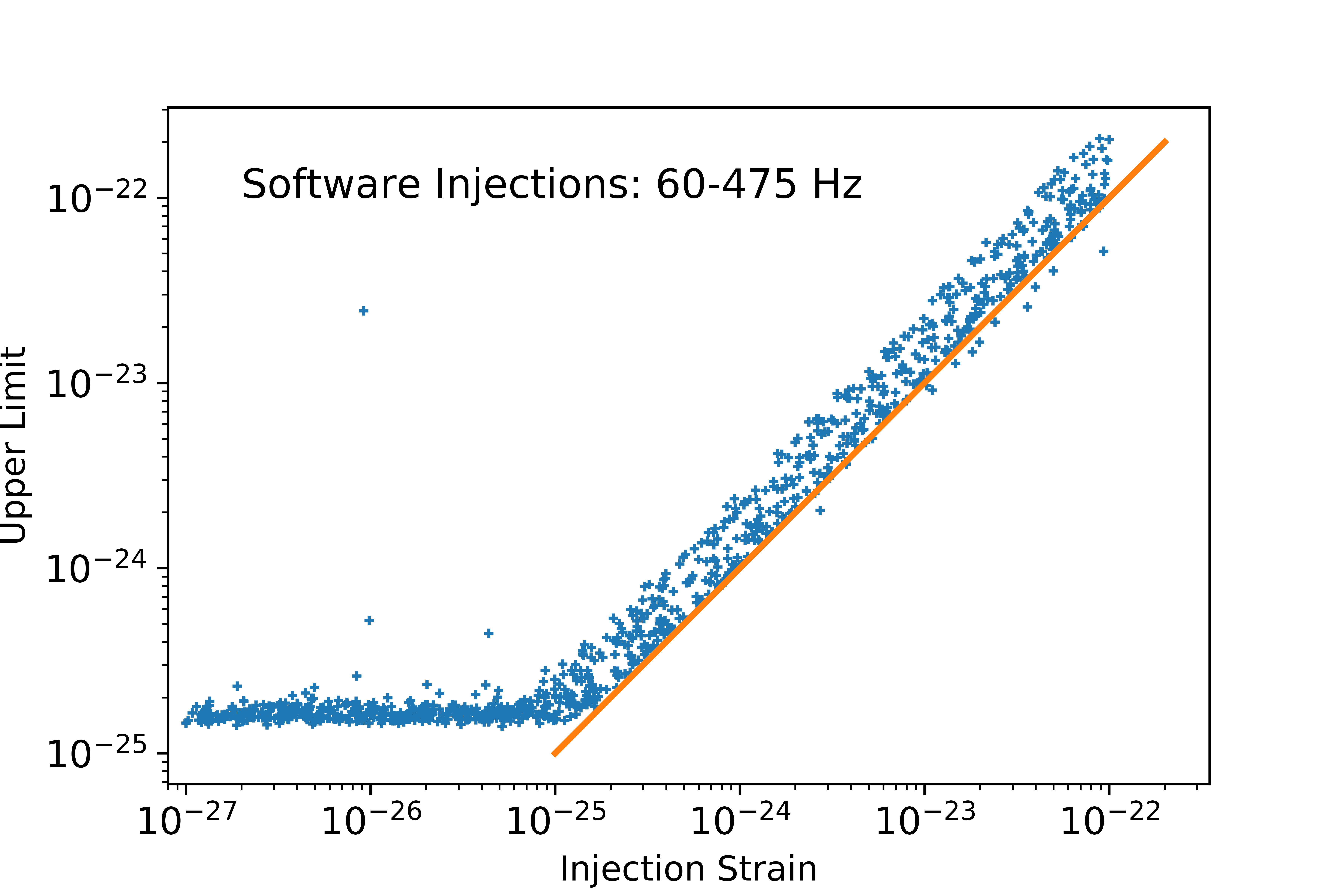}
    \includegraphics[width=3.3in]{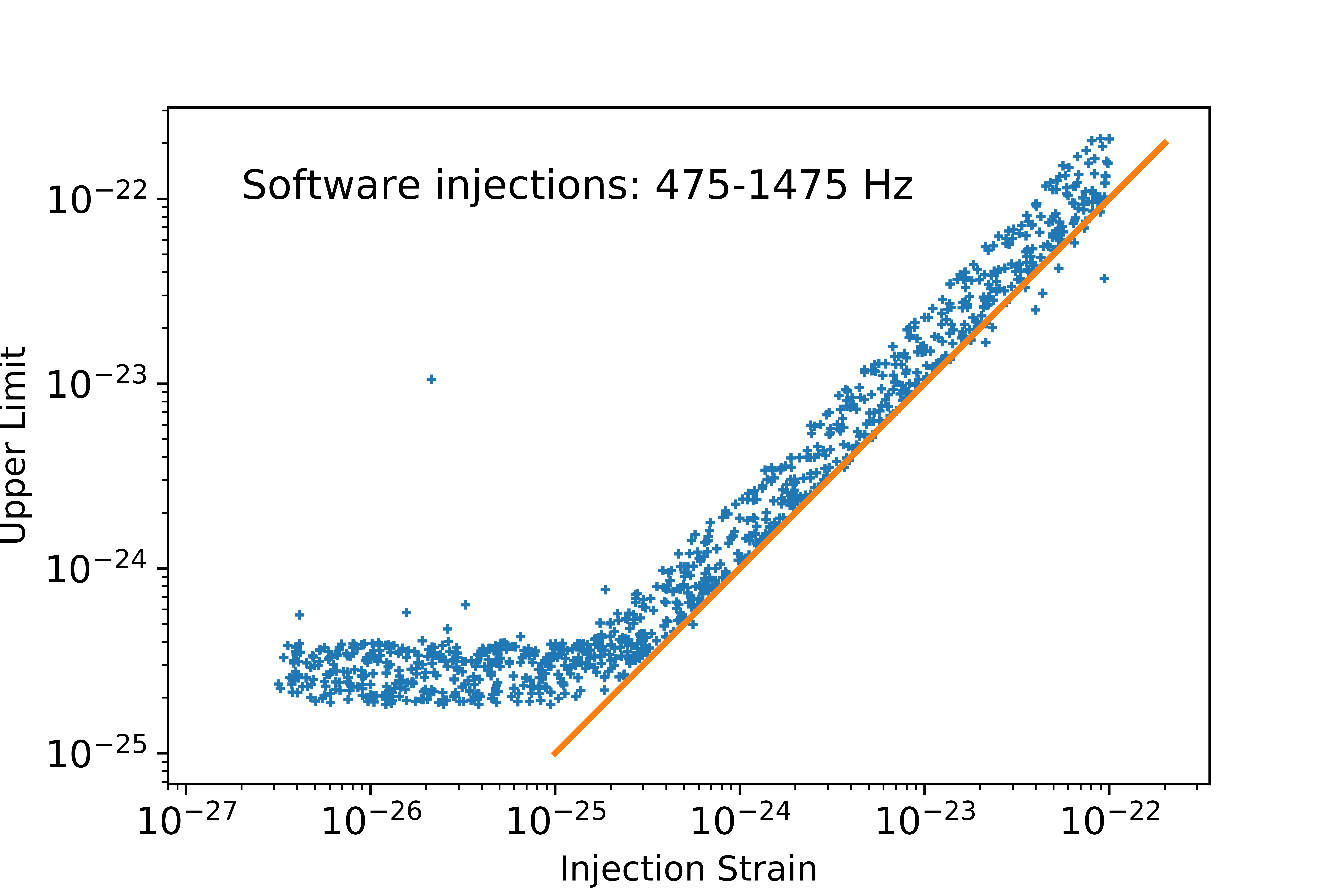}
    \includegraphics[width=3.3in]{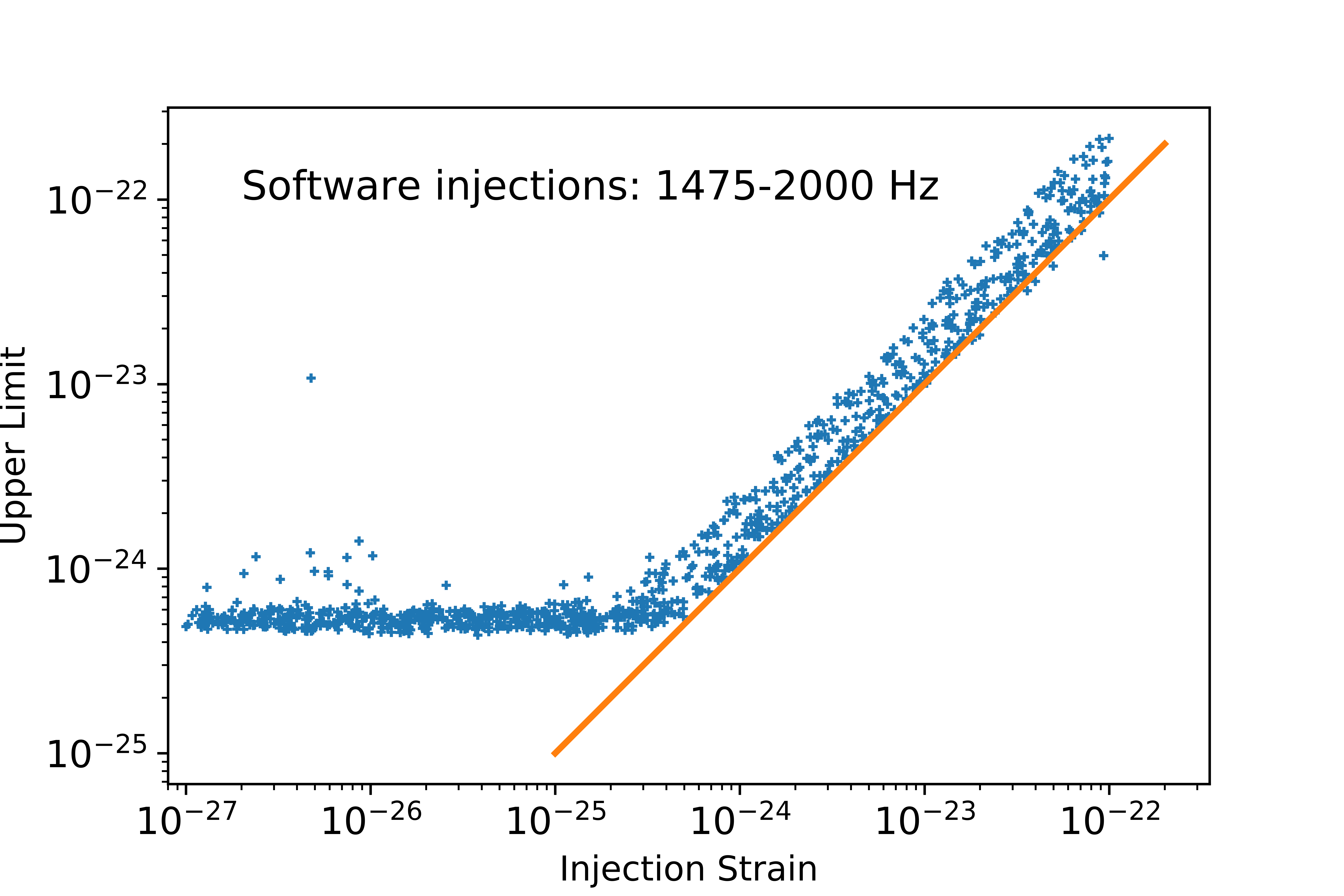}
    \fi
    \caption[Upper limit versus injected strain]{Illustration of \pf\ upper limit validation for very low, low, mid and high frequencies.
      Each point in the upper left, upper right, lower left and lower right panels represents a separate
      injection in the 20-60\,Hz, 60-475\,Hz, 475-1475\,Hz and 1475-2000\,Hz frequency ranges, respectively.
      Each established upper limit (vertical axis) is compared against the
      injected strain value defined by the horizontal axis (diagonal line defines equality of upper limit obtained and
      true injected strain). These injections are for random polarizations. Corresponding scatterplots for linearly
      polarized injections confirm the desired 95\%\ coverage, regardless
      of polarization.}
    \label{fig:ul_vs_strain}
  \end{center}
\end{figure*}

To allow robust analysis of the entire spectrum, including regions with severe spectral artifacts, a
{\em Universal statistic} algorithm~\cite{bib:universal_statistic,bib:O1allsky1} is used for establishing upper limits.
The algorithm is derived from the Markov inequality and shares its independence from the underlying noise distribution.
It produces upper limits less than $5$\% above optimal in case of Gaussian noise. In non-Gaussian bands,
it can report values larger than what would be obtained if the true underlying
distribution were known, but the upper limits are always
at least 95\% valid. Figure~\ref{fig:ul_vs_strain} shows results of a high-statistics
``software injections'' simulation run performed as described in~\cite{bib:S5allsky2}.
Correctly established upper limits lie above the dashed diagonal lines (defining equality between upper limit obtained and
true injection strain) in each panel, corresponding to
four selected sub-bands [SFT coherence times]: 20-60 Hz [7200s], 60-475 Hz [7200s], 475-1475 Hz [3600s] and 1475-2000 Hz [1800s].
Performance for the 7200s-SFT 20-60 Hz and 60-475 Hz bands are shown separately because of the proliferation
of spectral line artifacts below 60 Hz, primarily in the H1 data. The breakpoint frequencies of 475 Hz and 1475 Hz for
decreasing SFT coherence time (Table~\ref{tab:sft_info}) are those used in the O1 PowerFlux
search~\cite{bib:O1allsky1,bib:O1allsky2}, marking the starts of bands disturbed by 1st and 3rd violin mode harmonics.
Additional band-specific parameters for the initial stage of the search are listed in Table~\ref{tab:tuning_info}.

\begin{table*}[htb]
\begin{center}
  \begin{tabular}{lcccc}\hline
    \T\B                                         &  20-475 Hz      & 475-1475 Hz       & 1475-2000 Hz \\
    \hline\hline
    \T\B Frequency bin width                        &  0.139 mHz       & 0.278 mHz          & 0.556 mHz  \\
    \hline
    \T\B Number of spin-down templates\strut        &   66            &  34        &  18     \\
    \hline
    \T\B Spin-down step size (Hz/s)\strut        &  $\sci{1.692}{-10}$ & $\sci{3.333}{-10}$  & $\sci{6.667}{-10}$ \strut        \\
    \hline
     \T\B H1/L1 frequency mismatch tolerance (mHz)  &   2.5          &   2.5             & 2.5         \\     
    \hline
     \T\B H1/L1 spin-down mismatch tolerance (Hz/s)   &  $\sci{3.0}{-10}$ &   $\sci{3.0}{-10}$    & $\sci{3.0}{-10}$        \\     
    \hline
\end{tabular}
  \caption{Information on the band-specific parameters for the initial search (stage 0) and outlier follow-up stages.
    The number of spin-down steps and the spin-down step size refer to the templates used in the stage-0 search.
    The tolerances refer to the consistency requirements between H1 and L1 outliers used to define candidates selected
    for stage-1 follow-up. Other search parameters relevant to all search bands are described in the text.}
\label{tab:tuning_info}
\end{center}
\end{table*}

\begin{table*}[htbp]
\begin{center}
\begin{tabular}{llccccc}\hline
\T\B Stage & Instrument sum & {Phase coherence $\delta$} & \multicolumn{1}{c}{Spin-down step} & \multicolumn{1}{c}{Sky refinement} & \multicolumn{1}{c}{Frequency refinement} & \multicolumn{1}{c}{SNR increase} \\
 & & \multicolumn{1}{c}{rad} & \multicolumn{1}{c}{Hz/s} & $\alpha\times\delta_0$  &  & \multicolumn{1}{c}{\%}\\
\hline \hline
\multicolumn{7}{c}{\T\B 20-60\,Hz frequency range, 7200\,s SFTs, 0.0625\,Hz frequency bands} \\
  0 & Initial/upper limit incoherent & NA      & $\sci{1.692}{-10}$ & $1$   & $1/2$ & -- \\
  1 & incoherent                        & $\pi/2$ & $\sci{1.0}{-10}$   & $1/4\times1/4$ & $1/8$ & 20 \\
  2 & coherent                          & $\pi/2$ & $\sci{5.0}{-11}$   & $1/4\times1/4$ & $1/8$ & 10  \\
 \hline
\multicolumn{7}{c}{\T\B 60-475\,Hz frequency range, 7200\,s SFTs, 0.0625\,Hz frequency bands} \\
  0 & Initial/upper limit incoherent & NA      & $\sci{1.692}{-10}$ & $1$   & $1/2$ & -- \\
  1 & incoherent                        & $\pi/2$ & $\sci{1.692}{-10}$ & $1/4\times1/4$ & $1/8$ & 20 \\
  2 & coherent                          & $\pi/2$ & $\sci{5.0}{-11}$   & $1/4\times1/4$ & $1/8$ & 10  \\
 \hline
  \multicolumn{7}{c}{\T\B 475-1475\,Hz frequency range, 3600\,s SFTs, 0.125\,Hz frequency bands} \\
  0 & Initial/upper limit incoherent & NA      & $\sci{3.33}{-10}$  & $1$   & $1/2$ & -- \\
  1 & incoherent                        & $\pi/2$ & $\sci{3.33}{-10}$  & $1/4\times1/4$ & $1/8$ & 15 \\
  2 & coherent                          & $\pi/2$ & $\sci{5.0}{-11}$   & $1/4\times1/4$ & $1/8$ & 10  \\
 \hline
 \multicolumn{7}{c}{\T\B 1475-2000\,Hz frequency range, 1800\,s SFTs, 0.25\,Hz frequency bands} \\
  0 & Initial/upper limit incoherent & NA      & $\sci{6.67}{-10}$  & $1$   & $1/2$ & -- \\
  1 & incoherent                        & $\pi/2$ & $\sci{3.33}{-10}$  & $1/4\times1/4$ & $1/8$ & 15 \\
  2 & coherent                          & $\pi/2$ & $\sci{5.0}{-11}$   & $1/4\times1/4$ & $1/8$ & 10  \\
 \hline
\end{tabular}
\caption[Outlier follow-up parameters]{\pf\ outlier follow-up parameters. Stage 1 and higher use a loose-coherence algorithm for demodulation. The sky (both right ascension $\alpha$ and declination $\delta_0$) and frequency refinement parameters are relative to values used in the stage 0 \pf\ search}
\label{tab:followup_parameters}
\end{center}
\end{table*}

\begin{figure*}[htbp]
  \begin{center}
    \ifshowfigs
    \includegraphics[width=7.5in]{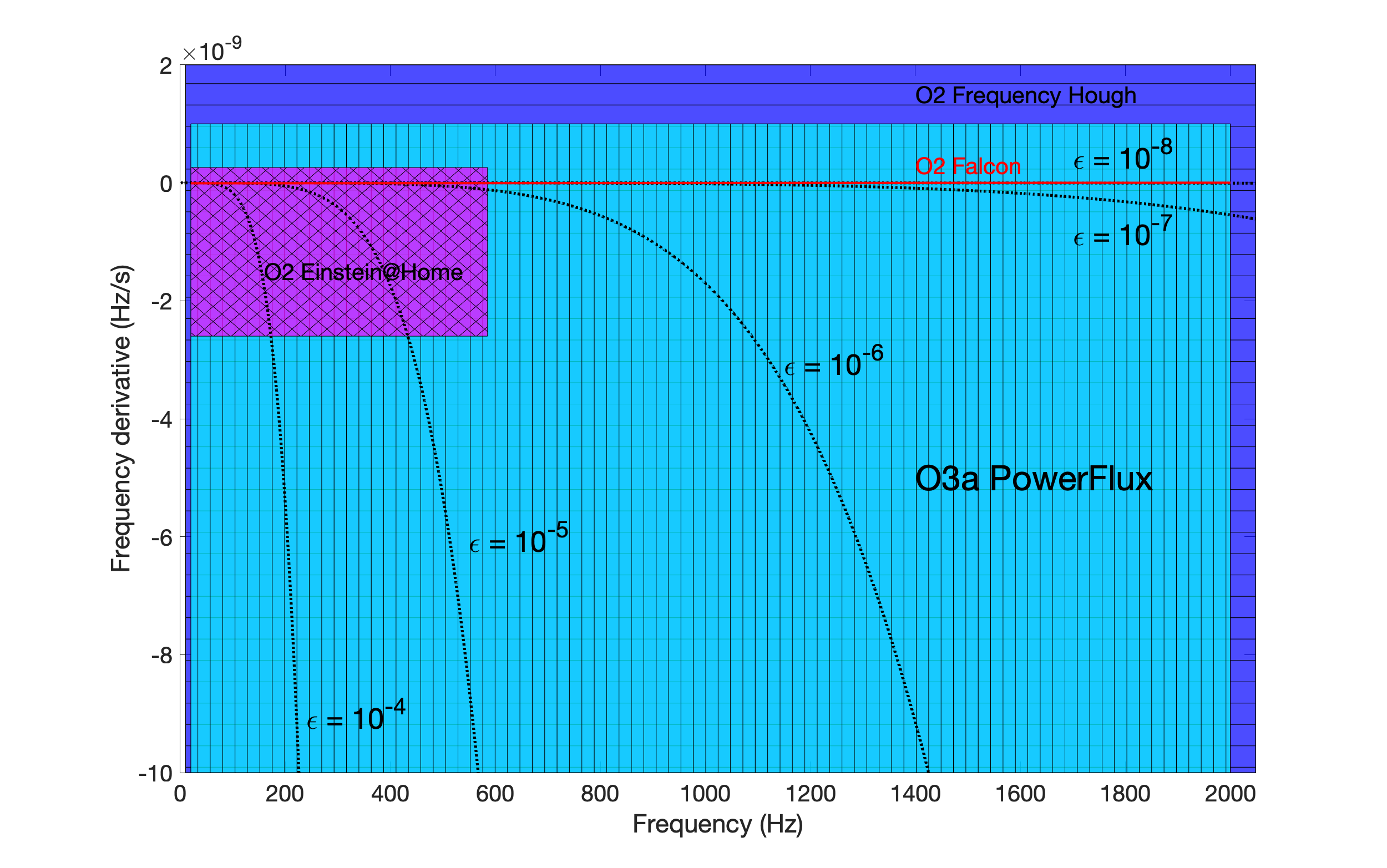}
    \fi
    \caption[Comparison of parameter space coverage]{Comparison in frequency and spin-down for this search with those of previous
      all-sky searches of LIGO O2 data. The shaded rectangle with vertical bars shows the 20--2000 Hz and
      $-10^{-8}$--$10^{-9}$ Hz/s range for this O3a search. The slightly larger rectangle with horizontal bars shows the
      region searched in the O2 data with the Frequency Hough method~\cite{bib:O2allsky,bib:O2allskyRome}. The smaller rectangle
      with crossed diagonal bars shows the region searched by the distributed-computing project Einstein@Home~\cite{bib:O2EatH}.
      The solid line at zero spin-down depicts the specialized O2 search for low-ellipticity millisecond pulsars using the
      Falcon method~\cite{bib:O2Falcon1,bib:O2Falcon2,bib:O2Falcon3} (the thickness of the line overstates the coverage in spin-down range).
      The dotted curves indicate contours of constant equatorial ellipticity $\epsilon$ = ($10^{-8}$, $10^{-7}$, $10^{-6}$, $10^{-5}$ and $10^{-4}$) for a star with stellar spin-down dominated by gravitational wave emission.}
    \label{fig:parameterspace}
  \end{center}
\end{figure*}

\subsection{Outlier follow-up}
\label{sec:outlier_followup}
  
A follow-up search for detection is carried out for high-SNR outliers found in stage 0.
The outliers are subject to an initial coincidence test.  For each
outlier with $\SNR>7$ in the combined H1 and L1 data, we require there
to be outliers in the individual detector data of the same small sky patch,
approximately square with side length $\sim$30 mrad $\times$ (100 Hz / frequency) that have $\SNR>5$ and
match the parameters of the combined-detector outlier within 2.5~mHz in frequency and
$\sci{3}{-10}$\,Hz/s in spin-down.  The combined-detector SNR is additionally required to be above both
single-detector SNRs, in order to suppress single-detector instrumental artifacts, except for unusually
loud outliers (combined, H1 and L1 SNRs all greater than 20).

The identified outliers using combined data are then passed to the follow-up stage using a loose-coherence
algorithm~\cite{bib:loosecoherence1} with progressively 
improved determination of frequency, spin-down, and sky location.

As the initial stage 0 sums only powers, it does not use the relative phase between interferometers,
which results in some degeneracy among sky position, frequency, and spin-down. The first loose-coherence
follow-up stage (1) demands greater temporal
coherence (smaller $\delta$) within each interferometer, which should
boost the SNR of viable outliers, but combines H1 and L1 power sums incoherently,
The subsequent stage (2) uses combined H1 and L1 data coherently, providing tighter bounds on outlier location.

Testing of the stages 0--2 pipeline is performed for frequency bands searched via software injections
using the same follow-up procedure. The recovery criteria also require that an outlier close to the true injection location (within 2.5~mHz in frequency
$f$, $\sci{3}{-10}$~Hz/s in spin-down and 28.5~rad$\cdot$Hz$/f$ in sky location) be found and successfully pass
through all stages of the detection pipeline.

\begin{figure*}[htbp]
  \begin{center}
    \ifshowfigs
    \includegraphics[width=3.3in]{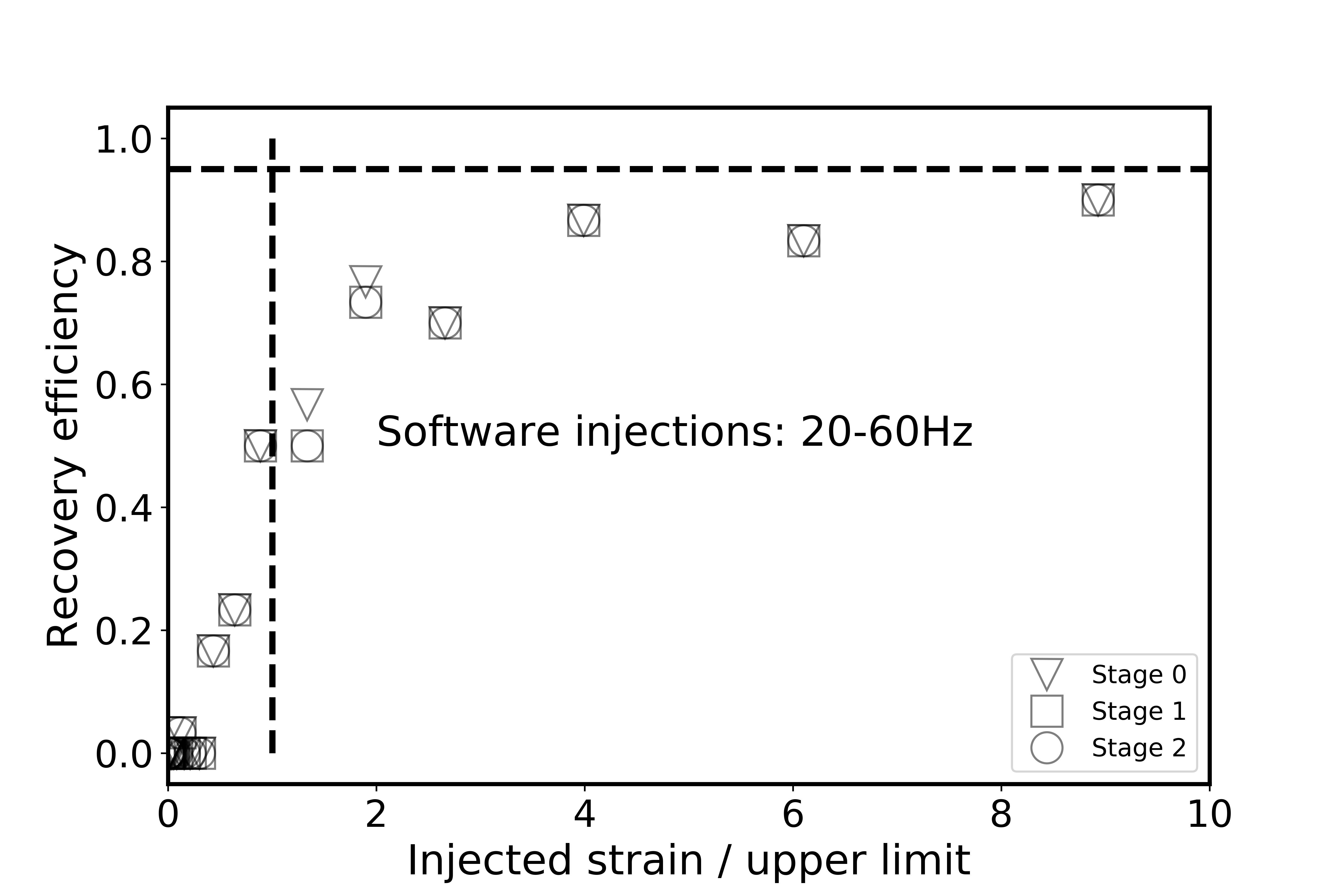}
    \includegraphics[width=3.3in]{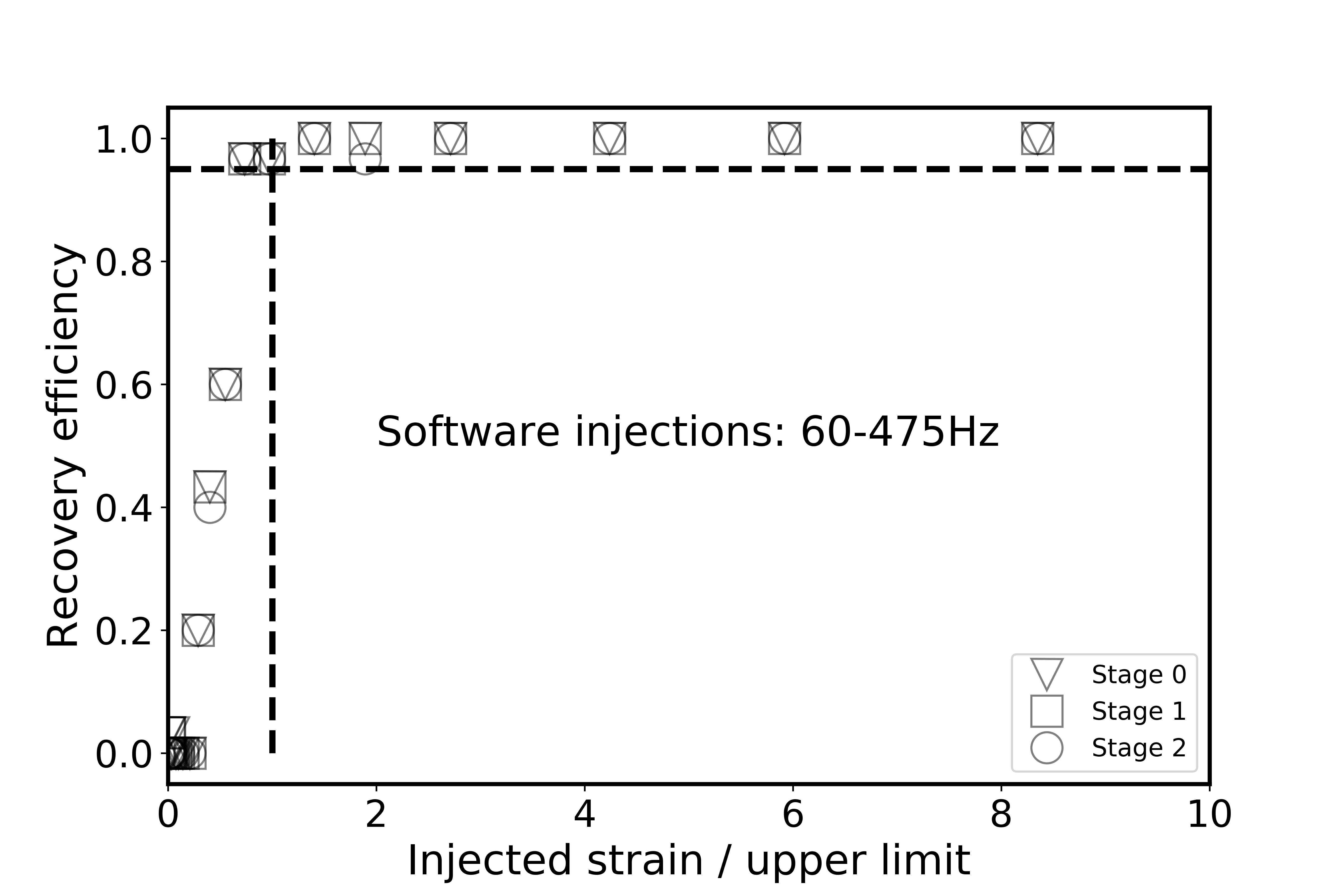}
    \includegraphics[width=3.3in]{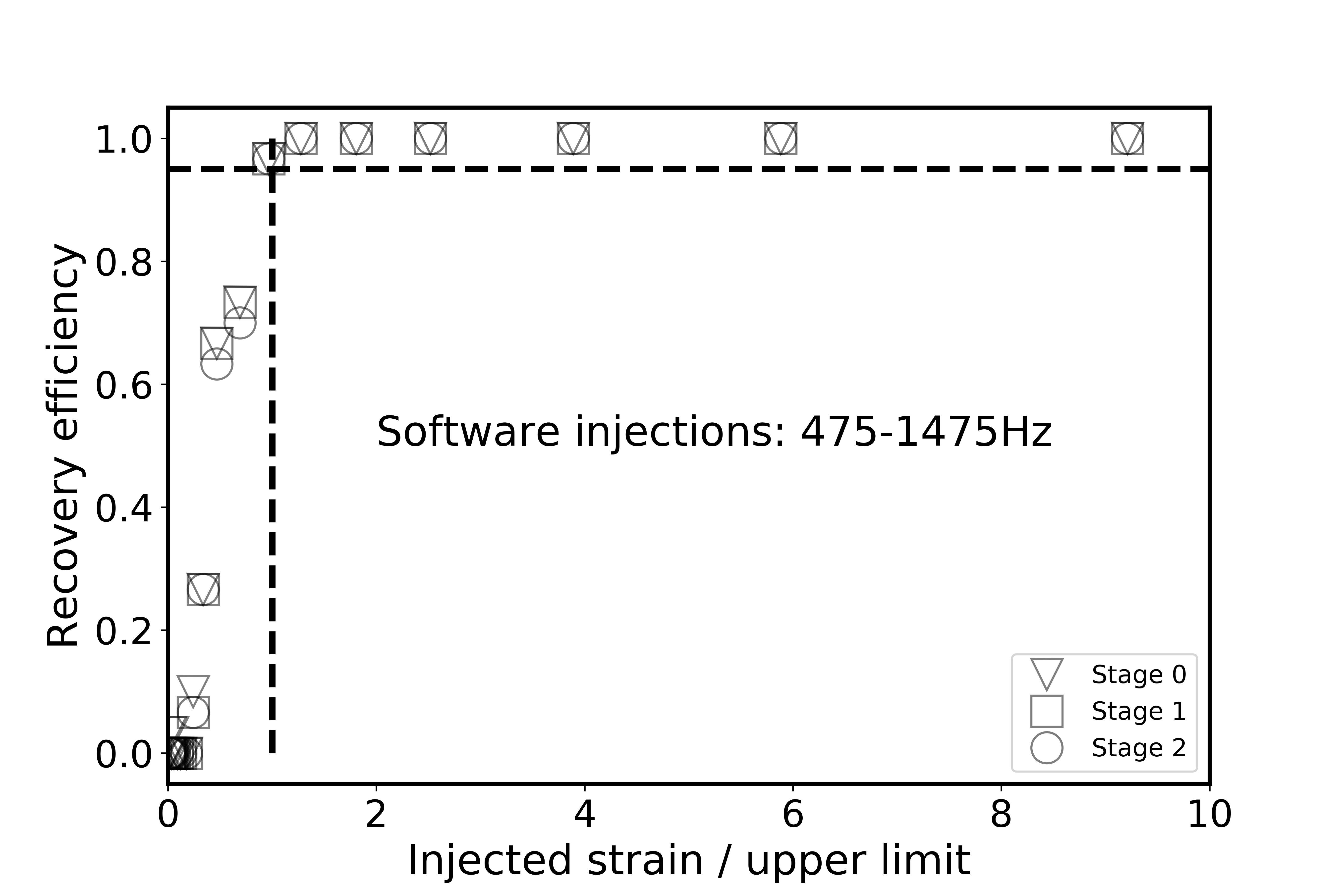}
    \includegraphics[width=3.3in]{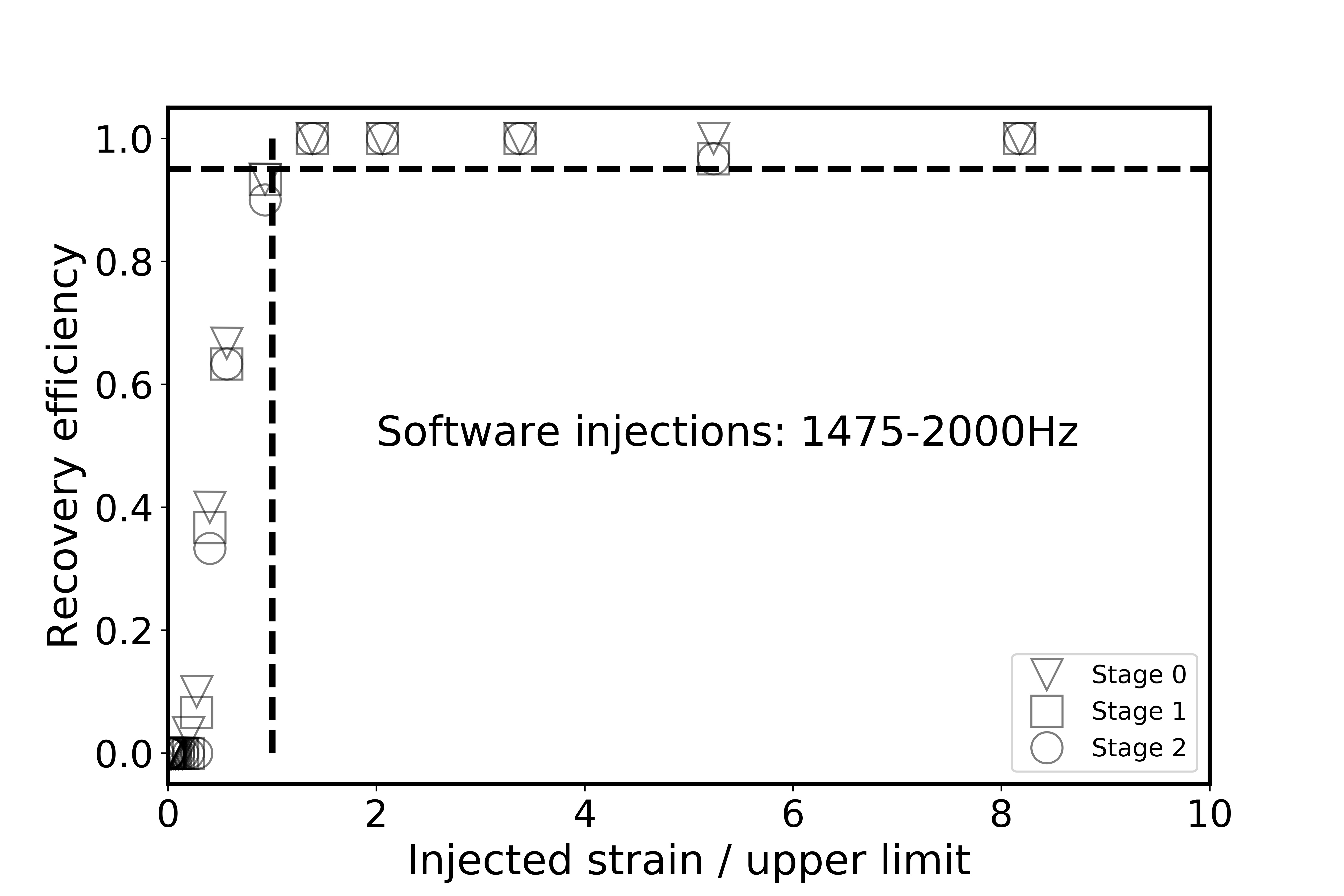}
    \fi
    \caption[Injection recovery]{Injection (software simulations) recovery efficiencies
      in the 20-60\,Hz, 60-475\,Hz, 475-1475\,Hz and 1475-2000\,Hz frequency bands are shown
      in the upper left, upper right, lower left and lower right panels, respectively, for stages 0, 1 and 2 of the search.
      The injected strain divided by the 95\%\ CL upper limit in its band (without injection)
      is shown on the horizontal axis.
      The percentage of surviving injections is shown on the vertical axis,
      with a horizontal dashed line drawn at the $95$\% level. The vertical dashed line
      marks a relative strain of unity. Ideally, the recovery efficiencies should lie
      to the left of the vertical line or above the horizonal line. One observes, however,
      that the ideal recovery efficiency is significantly degraded below 60 Hz, where large instrumental
      artifacts in the H1 data and only modest expected Doppler modulations make clean signal recovery more challenging.
      Most of the degradation stems from the band below 27 Hz for which H1 line contamination is severe and for
      which the H1 noise floor is substantially higher than the L1 noise floor.}
   \label{fig:injection_recovery}
  \end{center}
\end{figure*}

Injection recovery efficiencies from simulations covering the major sub-bands (20-475 Hz, 475-1475 Hz, 1475-2000 Hz) are
shown in Fig.~\ref{fig:injection_recovery} for stages 0, 1 and 2, which confirm that 95\%\ signal
recovery is comparable to the 95\%\ upper limit in all bands, as desired, except for the region
below 60 Hz for which spectral line artifacts heavily contaminate the H1 data set.
Injections in vetoed frequency bands (see section~\ref{sec:outliers} and Table~\ref{tab:outlier_badbands}) are not included in these graphs.

As in previous \pf\ analyses with loose-coherence follow-up~\cite{bib:O1allsky1,bib:O1allsky2},
only a mild influence from parameter mismatch is expected, as the parameters are chosen to accommodate the worst
few percent of injections. The follow-up procedure establishes very wide margins for outlier follow-up.
For example, when transitioning from the semi-coherent stage 0 to the loose-coherence stage 1 below 475 Hz, the effective
coherence length increases by a factor of 4. The average true signal SNR should then increase by more than $40$\%.
But the threshold used in follow-up is only 15--20\%, depending on frequency, which accommodates unfavorable noise conditions,
template mismatch, and detector artifacts.

Although a similar prior analysis~\cite{bib:O1allsky1,bib:O1allsky2} of O1 data used additional stages (3, 4) of
loose coherence, we choose after stage 2 to inspect candidates manually. The small number not found to be contaminated by
obvious instrumental artifacts are followed up in the full O3 data set by
a search similar to the \pf\ O3a stage 1, but using much finer spin-down stepping of $\sci{1}{-11}$ Hz/s, and refinement factors of 1/4
for both sky (right ascension and declination each) and frequency stepping, to exploit the improved SNR and parameter
resolution possible in the nearly doubled observation span. 
Any outlier surviving this full-O3 follow-up is explored via
a more sensitive method, implemented in \pyfstat~\cite{bib:PyFstatMethod1,bib:PyFstatZenodo} which
uses Markov Chain Monte Carlo (MCMC) techniques~\cite{bib:AshtonPrix,bib:PyFstatMethod2}
to explore small regions of parameter space in successive stages of increasing coherence times.

The full-O3 \pf\ follow-up search uses nearly 11 months of
data (compared to the 6-month O3a period), leading to an expected increase in SNR by a factor of approximately $\sqrt{11/6} = 1.35$,
with an additional $\sim$40\%\ increase from the use of stage-1 loose coherence, leading to a nominal expectation of SNR gain of
a factor of $\sim$1.9. Software simulations of signals with amplitudes as weak as 1/3 of the upper limit in a given band,
however, reveal large fluctuations above and below this expectation for all search bands.
These simulations suggest applying a conservative (negligible efficiency loss)
threshold on the SNR increase of only 20\%, which is nonetheless effective in suppressing most noise artifacts.

The final follow-up method applied to any survivors of the full-O3 loose-coherence check
uses the Python-based \pyfstat~\cite{bib:PyFstatMethod1,bib:PyFstatZenodo} software
infrastructure to combine a
MCMC approach~\cite{bib:AshtonPrix,bib:PyFstatMethod2} with semi-coherent summing of the well known \Fstat\ detection
statistic~\cite{bib:JKS}. In this approach, the parameter space near to those values from a stage-2 outlier is sampled
randomly according to a certain probability density function determined by the \Fstat\ likelihood function.
We closely follow an implementation~\cite{bib:PyFstatMethod2} applied to a recent analysis of O2 data to follow up
on outliers emerging from several prior O2 continuous wave searches.

Briefly, the O3a observation time is divided into \nseg\ segments, for each of which the \Fstat\ is computed
over a coherence time approximately equal to the observation time divided by \nseg. For each point sampled
in parameter space, the sum of the \Fstat\ values is computed to form a total detection statistic.
This procedure is repeated, decreasing the number of segments (increasing the coherence time),
using the resulting MCMC-maximized  \Fstat\ sum as the seed for the next stage, with a
consequent reduction in parameter space volume searched.
For this O3a analysis we choose five successive stages of follow-up with decreasing values of \nseg\ = 500 (coherence
time of 0.36 day), 250, 55, 5 and 1. A random sample of 600 off-source sky locations having the same declination as the putative
signal direction, but separated by more than 90 deg from that direction, is used to determine a non-signal expectation for
the background distributions in the same frequency band~\cite{bib:PyFstatMethod2}.
A Bayes factor $B_{\rm SN}$ is computed from the change in
\Fstat\ values for a nominal signal compared to the  empirical background  distribution in the last stage (\nseg=5 to \nseg=1).

\subsection{Hardware injections}
\label{sec:hwinjections}

During the O3 run 18 hardware injections were used to simulate
particular CW signals, as part of detector response validation, including long-term
phase fidelity. The injections were imposed via radiation pressure from auxiliary lasers~\cite{bib:O1hwinjections}.
For reference, Table~\ref{tab:hwinjections} lists the key source parameters for the
14 injections relevant to this analysis (labeled Inj0-Inj12 and Inj14), namely those that simulate isolated
neutron stars with nominal frequencies between 20 and 2000 Hz. 
In general, the injection amplitudes used in the O3 run are substantially lower than those used in
previous observing runs. For this reason, some injections with identical source parameters, except for stronger signal amplitudes,
were detected with high SNR in the O2 run, but are not detected in this analysis,
although they can be found with targeted matched-filter analyses.

For each injection in Table~\ref{tab:hwinjections}, the column labeled ``UL sig bin''
gives the 95\%\ UL for the corresponding band. Ideally,
that value should exceed the true injected amplitude $h_0$ for at least 95\%\ of the injections.
In this case, that statement holds for all 14 injections. The column labeled ``UL ctrl bins'' shows the average
95\%\ UL for the six nearest neighboring frequency subbands as a rough guide to the expected value in
the absence of an injection (or true signal). 

The last column states whether or not the injection survives all stages of loose-coherence follow-up.
As hoped, all five injections with a true $h_0$ amplitude above the expected background estimation of the
UL do survive these follow-up stages, as do two additional injections with somewhat smaller amplitudes.

\begin{table*}[htbp]
\begin{center}
\begin{tabular}{lD{.}{.}{6}rD{.}{.}{5}D{.}{.}{5}rrrc}
\hline
Label & \multicolumn{1}{c}{Frequency} & \multicolumn{1}{c}{Spindown} & \multicolumn{1}{c}{$\RAJ$} & \multicolumn{1}{c}{$\DECJ$} & \multicolumn{1}{c}{$h_0$} & \multicolumn{1}{c}{UL} & \multicolumn{1}{c}{UL} & \multicolumn{1}{c}{Detected?}\\
 & \multicolumn{1}{c}{Hz} & \multicolumn{1}{c}{nHz/s} & \multicolumn{1}{c}{degrees} & \multicolumn{1}{c}{degrees} & \multicolumn{1}{c}{true} & \multicolumn{1}{c}{sig bin}  & \multicolumn{1}{c}{ctrl bins} & \\
\hline \hline
Inj0   &  265.575086  & $\sci{-4.15}{-3}$   &   71.55193     &  -56.21749 & $\sci{6.12}{-26}$ & $\sci{1.8}{-25}$ & $\sci{1.8}{-25}$ & {\color{blue}Yes} \\
Inj1   &  848.937350  & $\sci{-3.00}{-1}$   &   37.39385     &  -29.45246 & $\sci{5.47}{-25}$ & $\sci{8.1}{-25}$ & $\sci{3.6}{-25}$ & {\color{blue}Yes}\\
Inj2   &  575.163506  & $\sci{-1.37}{-4}$   &  215.25617     &    3.44399 & $\sci{7.59}{-26}$ & $\sci{2.7}{-25}$ & $\sci{2.7}{-25}$ & No\\
Inj3   &  108.857159  & $\sci{-1.46}{-8}$   &  178.37257     &  -33.4366  & $\sci{1.30}{-25}$ & $\sci{1.9}{-25}$ & $\sci{1.9}{-25}$ & No\\
Inj4   & 1390.806586  & $\sci{-2.54}{+1}$   &  279.98768     &  -12.4666  & $\sci{1.07}{-24}$ & $\sci{4.9}{-25}$ & $\sci{4.9}{-25}$ & {\color{blue}Yes\footnote{True spin-down value outside of nominal search range. Injection is recovered when search range is extended for this band.}}\\
Inj5   & 52.808324  & $\sci{-4.03}{-9}$     &  302.62664     &  -83.83914 & $\sci{3.99}{-25}$ & $\sci{5.7}{-25}$ & $\sci{3.2}{-25}$ & {\color{blue}Yes}\\
Inj6   &  145.444975 & $\sci{-6.73}{0\mathrm{\;\;\;}}$     &  358.75095     &  -65.42262 & $\sci{3.84}{-25}$ & $\sci{4.3}{-25}$ & $\sci{2.2}{-25}$ & {\color{blue}Yes}\\
Inj7   & 1220.434717 & $\sci{-1.12}{0\mathrm{\;\;\;}}$     &  223.42562     &  -20.45063 & $\sci{1.65}{-25}$ & $\sci{4.5}{-25}$ & $\sci{4.4}{-25}$ & No\\
Inj8   &  190.100215 & $\sci{-8.65}{0\mathrm{\;\;\;}}$     &  351.38958     &  -33.41852 & $\sci{1.30}{-25}$ & $\sci{1.8}{-25}$ & $\sci{1.7}{-25}$ & {\color{blue}Yes} \\
Inj9   &  763.847316 & $\sci{-1.45}{-8}$    &  198.88558     &   75.68959 & $\sci{1.30}{-25}$ & $\sci{3.1}{-25}$ & $\sci{3.0}{-25}$ & No\\
Inj10  &   26.332768 & $\sci{-8.50}{-2}$    &  221.55565     &   42.87730 & $\sci{6.26}{-25}$ & $\sci{1.9}{-24}$ & $\sci{1.2}{-24}$ & No\\
Inj11  &   31.424704 & $\sci{-5.07}{-4}$    &  285.09733     &  -58.27209 & $\sci{3.17}{-25}$ & $\sci{9.4}{-25}$ & $\sci{7.8}{-25}$ & No\\
Inj12  &   37.805210 & $\sci{-6.25}{0\mathrm{\;\;\;}}$     &  331.85267     &  -16.97288 & $\sci{2.63}{-25}$ & $\sci{4.8}{-25}$ & $\sci{4.9}{-25}$ & No\\
Inj14  & 1991.092292 & $\sci{-1.00}{-3}$     & 300.80284     &  -14.32394 & $\sci{1.83}{-24}$ & $\sci{2.1}{-24}$ & $\sci{8.1}{-25}$ & {\color{blue}Yes}\\
\hline
\end{tabular}\\
\caption[Parameters of hardware injections]{Key parameters of the hardware-injected simulated isolated-source continuous wave
  signals during the O3 data run (epoch GPS $1238166018$) with nominal frequencies
  between 20 and 2000 Hz, along with upper limits in the nominal signal bins and in the six nearest
  control bins. The last column states whether or not the injection is detected in outlier follow-up.
  Most of the injection amplitudes are significantly weaker than those used in prior observing runs,
  leading to fewer detections than in past analyses. The amplitudes quoted here are the {\it intended} values;
  uncertainties in the approximate actuation functions used in the online injection system may be as large
  as a few percent, depending on frequency. Also, a now-known error in the actuation function used for L1 injections
  led to an unintended time delay of approximately 122 $\mu$s, which corresponds to a phase lag in the highest-frequency
  injection (2991 Hz) of $\sim$130$^\circ$ and degrades coherent H1-L1 injection recovery.} 
\label{tab:hwinjections}
\end{center}
\end{table*}

\section{Search results}
\label{sec:results}

Carrying out the stage-0 analysis described above leads to a set of all-sky upper limits
(95\%\ CL) on strain amplitude for worst-case, linear polarization (relative stellar spin
orientation) and for best-case, circular polarization. That analysis also leads to an
initial set of outliers for follow-up with later analysis stages.
Whether or not a particular narrow frequency band contains an outlier,
the upper limits obtained remain valid.
As shown in Table~\ref{tab:hwinjections}, hardware injections were reliably recovered when
their injected strain was at least 0.7 times the recorded upper limit (and in one case as low as 0.34 times the upper limit).
The upper limits obtained from stage-0 analysis
are presented in section~\ref{sec:upperlimits}, along with corresponding astrophysical
sensitivities. Section~\ref{sec:outliers} presents the results of outlier follow-up.

\subsection{Upper limits and sensitivity}
\label{sec:upperlimits}

\begin{figure*}[htbp]
  \begin{center}
    \ifshowfigs
    \includegraphics[width=7.5in]{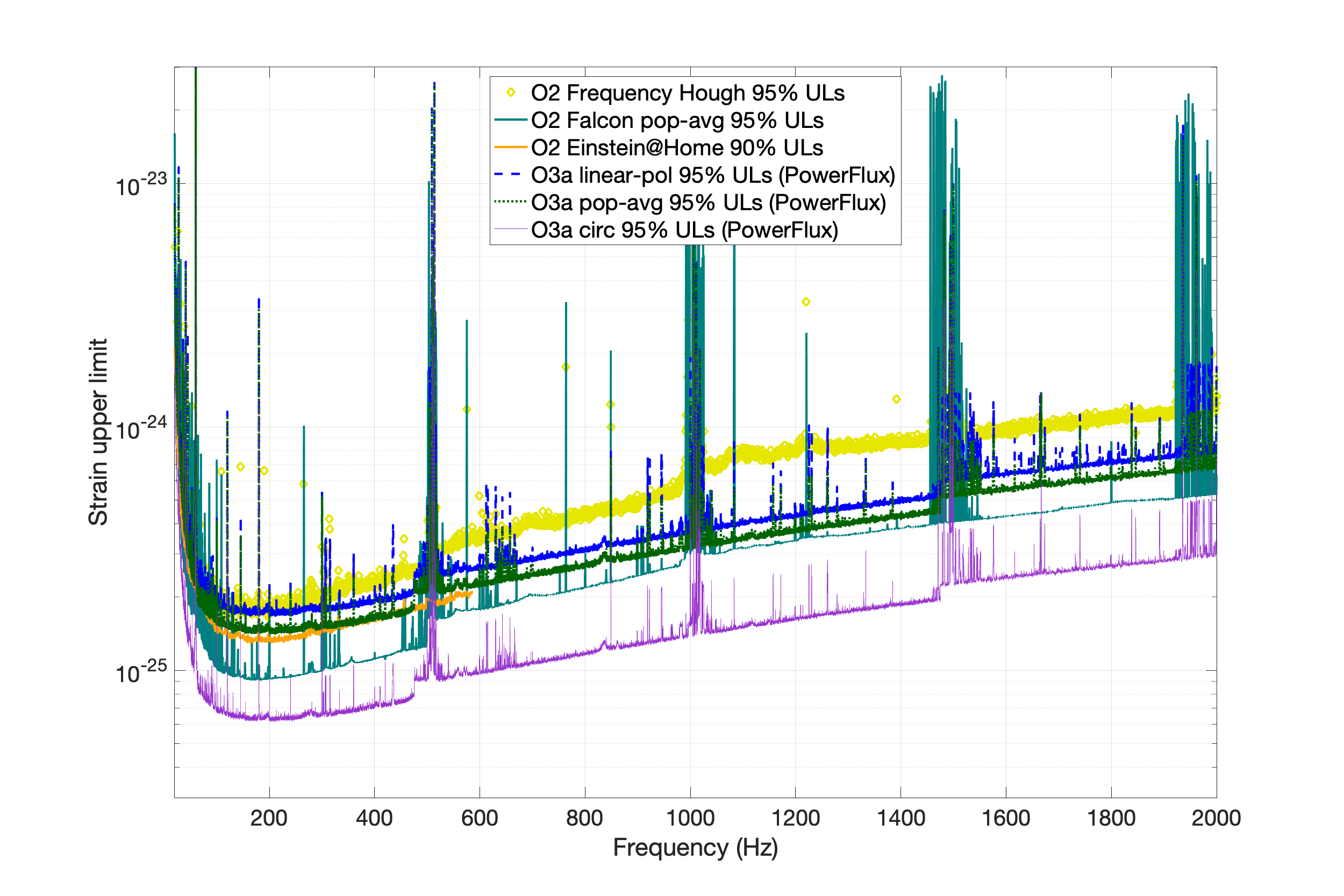}
    \fi
    \caption{Upper limits on gravitational strain amplitude for this O3a analysis and for previous O2 analyses.
      The dark blue curve shows worst-case (linearly polarized)
      $95$\%~CL upper limits in analyzed ($62.5$\,mHz, $125$\,mHz, $250$\,mHz) sub-bands for the
      broad (20-475 Hz, 475-1475 Hz, 1475-2000 Hz) search bands. The lowest, purple curve shows upper limits assuming a
      circularly polarized source. The middle (green) curve shows approximate
      population-averaged all-sky upper limits inferred from the circularly polarized limits.
      For comparison, the uppermost yellow-green diamonds show previous population-averaged 95\%\ CL all-sky upper limits
      derived from LIGO O2 data using the Frequency Hough method~\cite{bib:O2allsky,bib:O2allskyRome}.
      The orange curve shows 90\%\ CL upper limits from the O2 Einstein@Home search~\cite{bib:O2EatH}
      over the band 20-585.15 Hz and over a spin-down range that is 26\%\ of the range used here. 
      The light blue curve shows 95\%\ CL upper limits from the O2 Falcon search~\cite{bib:O2Falcon1,bib:O2Falcon2,bib:O2Falcon3}
      over the 20-2000 Hz band and a severely restricted spin-down range.
      See Fig.~\ref{fig:parameterspace} for a comparison
      of parameter space coverage from these various searchers.}
    \label{fig:upper_limits}
  \end{center}
\end{figure*}

Figure~\ref{fig:upper_limits} shows the upper limits obtained in this search as three curves, along with results from
other all-sky searches in the O2 data set~\cite{bib:O2allsky,bib:O2allskyRome,bib:O2EatH,bib:O2Falcon1,bib:O2Falcon2,bib:O2Falcon3}.
The upper (blue) curve shows the upper limits for a worst-case (linear) polarization. The lowest (red) curve shows
upper limits for an optimally oriented source (circular polarization). Both curves represent strict frequentist upper
limits on these worst-case and best-case orientations with respect to location on the sky, spin-down and frequency
within each narrow search band. For sensitivity comparisons with the previous O2 results,
the intermediate (green) curve represents approximate population-averaged upper limits
(over random sky locations and polarizations) and is
derived from the circular polarization curve with a simple scale factor (2.3), based on injection studies in test bands.

Each linear-polarization or circular-polarization point in Fig.~\ref{fig:upper_limits} represents a maximum over the sky,
except for a small excluded portion of the sky near the ecliptic poles,
which is highly susceptible to detector artifacts due to stationary frequency evolution produced by the combination of
frequency derivative and Doppler shifts~\cite{bib:S5allsky2}.

The O3a results presented here improve upon the previous O2 results in strain
sensitivity with factors ranging from $\sim$1.2 at signal frequencies of
$\sim$100 Hz to $\sim$2.0 at higher frequencies below the 1475-Hz breakpoint. The improvements at high frequencies come in part from
the improved detector noise achieved with quantum squeezing~\cite{bib:O3squeezing} and in part from using 
longer Fourier transform coherence times than were used in the similar O2 analyses.
At frequencies below $\sim$40 Hz, the O2 data was badly contaminated by
instrumental lines~\cite{bib:linespaper}, leading to poorer upper limits, and in some bands, precluding reliable upper limits
using the method of~\cite{bib:O2allskyRome}. The dramatically improved sensitivies obtained for low frequencies in this analysis come in part from
mitigation of many spectral lines between the O2 and O3 runs, especially in the L1 interferometer, and in part from the
use here of the ``universal statistic'' described in section~\ref{sec:upperlimit_determination}.

\begin{figure*}[htbp]
  \ifshowfigs
  \includegraphics[width=7.5in]{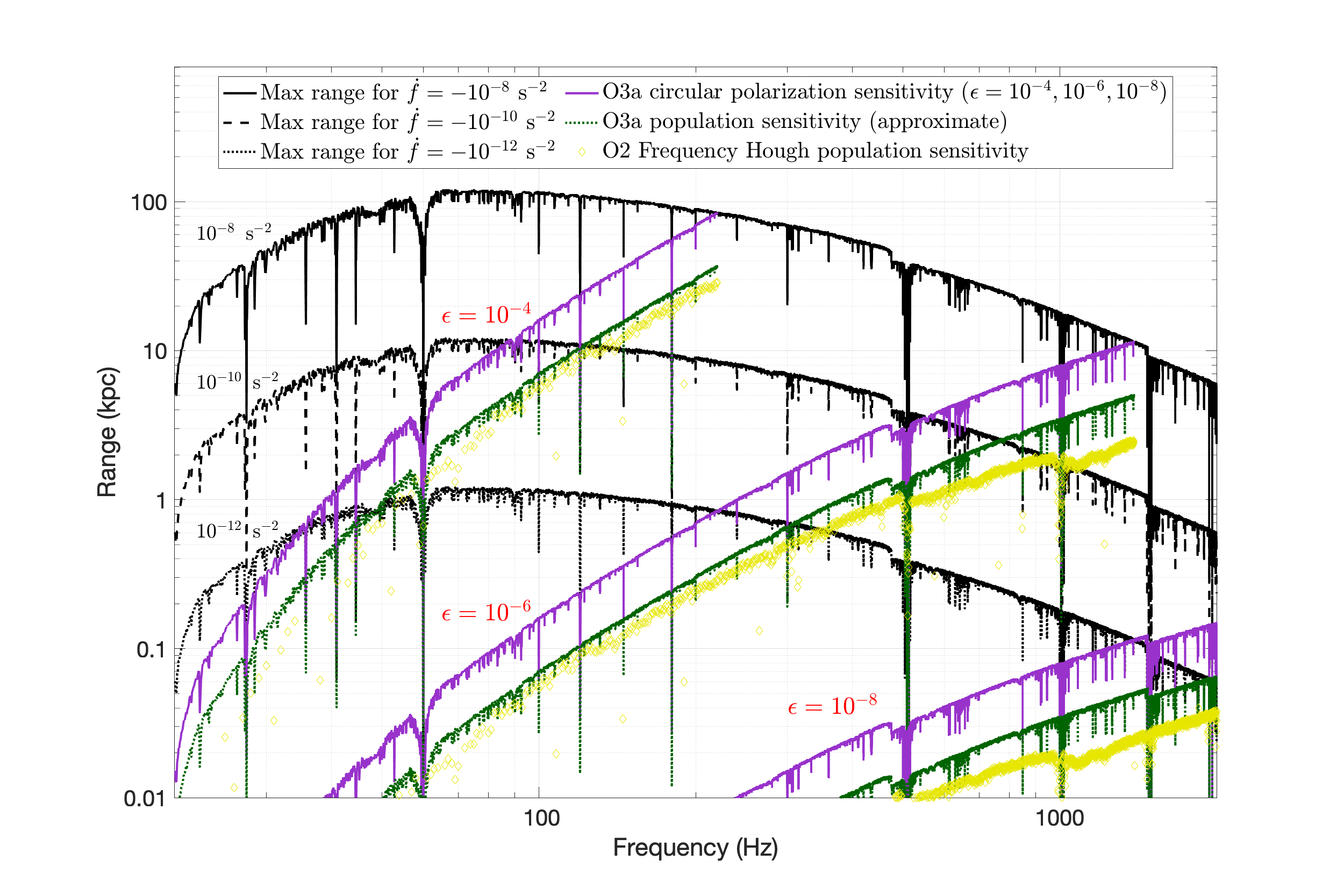}
  \fi
  \caption[Search range]{
    Ranges (kpc) of the \pf\ search for neutron stars
    spinning down solely due to gravitational radiation (``gravitars'') under different assumptions.
    The three sets of three curves (purple, green dotted, yellow-green diamonds)
    that generally rise with frequency are the ranges for which the O3a circular-polarization,
    O3a population-averaged and O2 population-averaged Frequency
    Hough~\cite{bib:O2allsky,bib:O2allskyRome} upper limits apply for three
    different assumed equatorial ellipticities of $\epsilon$ = $10^{-4}$, $10^{-6}$ and $10^{-8}$.
    In each case, the curve stops at the frequency at which the implied spindown magnitude
    at the corresponding range exceeds the maximum range allowed by the maximum spindown magnitude ($10^{-8}$ Hz/s)
    used in these analyses.
    The three black curves (solid, dashed, dot-dashed) that peak below 100 Hz represent the maximum possible ranges
    for which the O3a circular-polarization limits apply under the assumption that the
    maximum spindown magnitude is ($10^{-8}$, $10^{-10}$, $10^{-12}$) Hz/s.
    Searching for high spindown magnitudes simultaneously probes higher
    ellipticities and larger regions of the galaxy, reaching well beyond the galactic center ($\sim$ 8.5 kpc)
    at high ellipticities and low frequencies. More realistically, a neutron star's spin-down may be dominated
    by electromagnetic radiation, in which case searching for high spindown magnitude achieves a shorter search
    range, but enables detection of stars that would not be detected under the gravitar assumption.
    (color online)}
  \label{fig:range}
\end{figure*}

One can recast these upper limits on source amplitude as lower limits on the range at which
neutron stars with assumed equatorial ellipticity can reside.
Figure~\ref{fig:range} shows the ranges (kpc) of this search (circular-polarization and population-averaged)
\vs\ frequency for assumed equatorial ellipticity values
$\epsilon$ = $10^{-4}$, $10^{-6}$ and $10^{-8}$ and under the assumption that the {\it total} stellar spin-down magnitude does
not exceed the $10^{-8}$ Hz/s maximum used in this search.
The population-averaged ranges from prior O2 searches of comparable parameter space
coverage~\cite{bib:O2allsky,bib:O2allskyRome} are also shown for
these ellipticities. One sees a gain in range with respect to O2 analyses over the full frequency band,
with the greatest improvements seen at highest frequencies, as expected. The number of accessible neutron
stars is expected to rise at least quadratically as the range extends out into the plane of the galaxy and
approaches the denser regions near its center, at a distance of $\sim$8.5 kpc, enhancing the detection
probabilities from these range gains.

For reference, Fig.~\ref{fig:range} also shows this search's implied maximum ranges \vs\ frequency for a
``gravitar'' (star with spin-down dominated by gravitational radiation) for 
maximum spin-down magnitudes of $10^{-8}$, $10^{-10}$ and $10^{-12}$ Hz/s.
If a neutron star's spin-down has major contributions from 
electromagnetic radiation, the range to which this search is sensitive for fixed $\epsilon$ may be
reduced by the choice of maximum spin-down permitted.
Conversely, searching for spin-down magnitudes much higher than expected for a gravitar of given $\epsilon$
could allow detection of a star with more realistic emission contributions.

\subsection{Outliers}
\label{sec:outliers}

Outliers seen in the stage-0 search are followed up with loose-coherence stages of
increasing effective coherence time, as described in section~\ref{sec:outlier_followup},
with the SNR expected to increase for true signals. Table~\ref{tab:outlier_counts}
shows the counts of outliers seen at each stage for major sub-bands.
In the presence of Gaussian noise and no signal, one expects the outlier counts to
decrease monotonically with increasing stage as SNR increases are required for each advancement.
For particular sub-bands, however, one can see count increases from at least two contributions,
both associated with the finer sampling of parameter space in successive stages.
1) Hardware injections (section~\ref{sec:hwinjections}) from
a simulated signal naturally satisfy the SNR increase requirements demanded of a signal;
and 2) stationary line artifacts of finite bandwidth can be compatible with
signal templates having limited Doppler modulation or having partial cancellation
between seasonal modulation and assumed frequency derivative
for a certain region of the sky~\cite{bib:S4allsky}. 

\begin{figure*}[htbp]
  \ifshowfigs
  \includegraphics[width=6.5in]{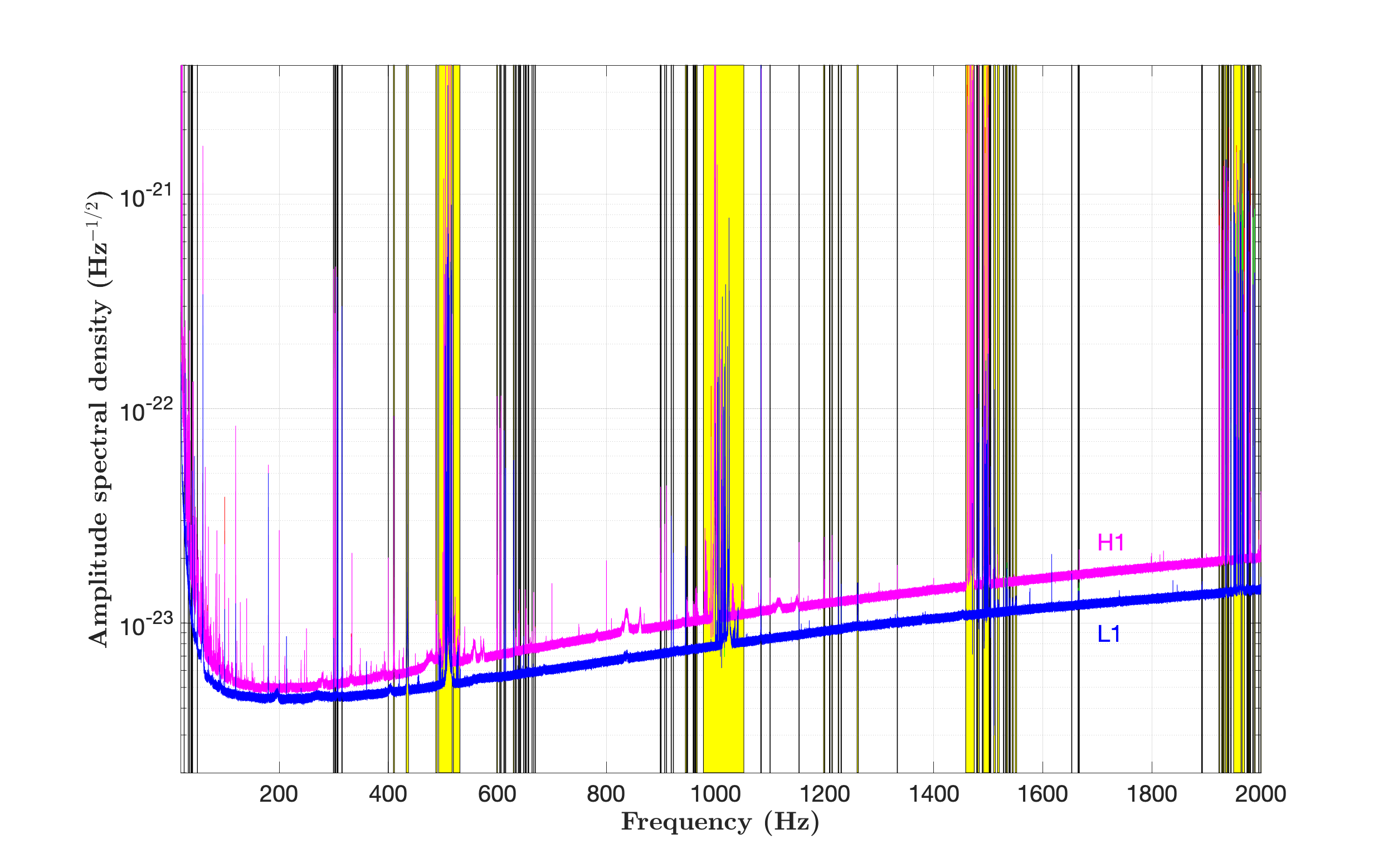}
  \includegraphics[width=6.5in]{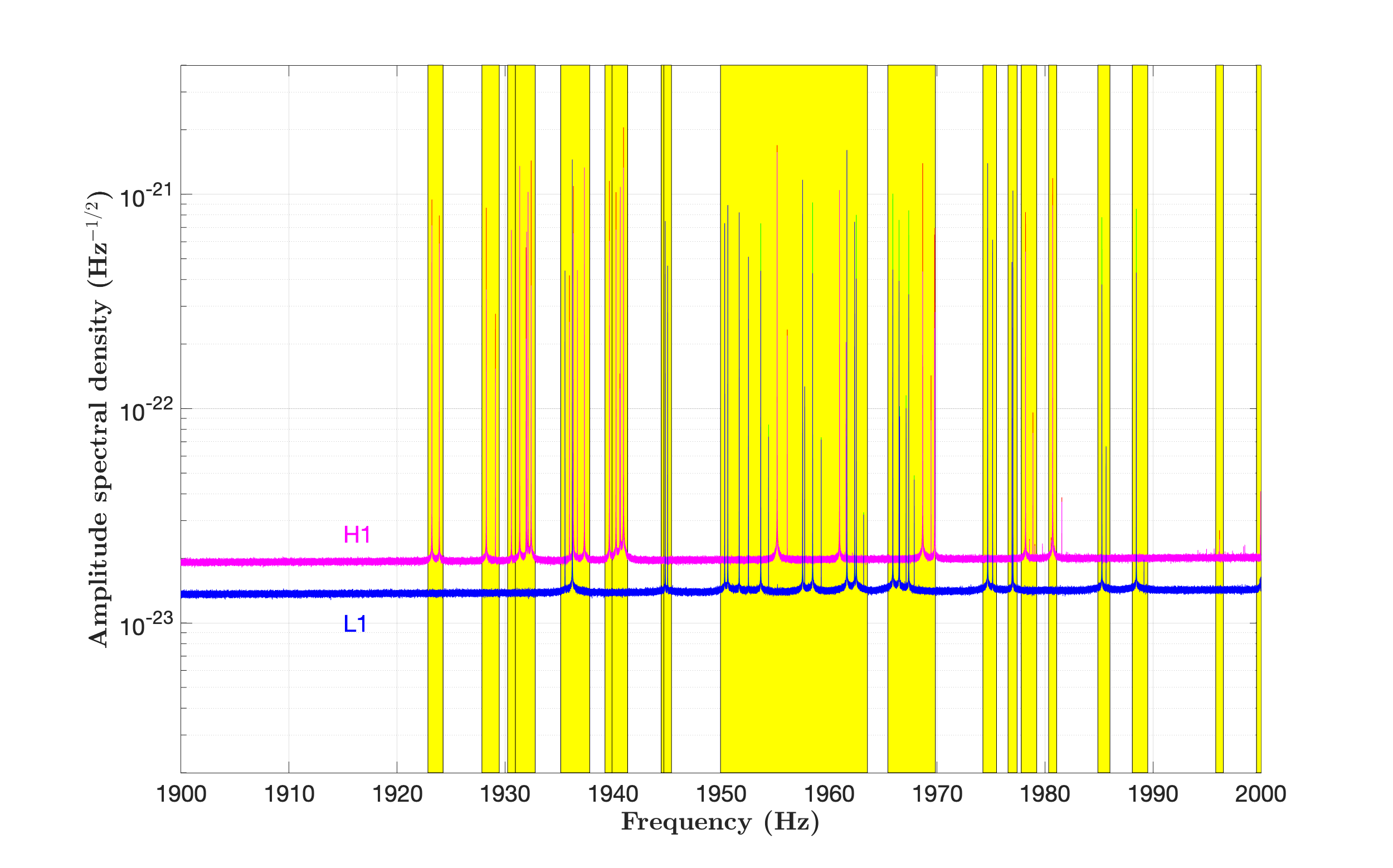}
  \fi
  \caption[Vetoed bands]{
    Shaded regions (yellow) overlaid on the run-averaged spectra (inverse-noise weighted) from the
    H1 (magenta) and L1 (blue) O3a data sets 
    indicate vetoed bands for which outlier follow-up is impeded by artifacts. The widest affected
    bands correspond to test-mass ``violin modes'' (ambient vibrations of the silica fibers from which LIGO mirrors are suspended)
    near multiples of 500 Hz.
    {\it Top panel:} Full 20-2000 Hz band. 
    {\it Bottom panel:} Magnification of example 1900-2000 Hz band, for which line artifacts
    are dominated by the 4th harmonics of violin modes. 
    (color online)}
  \label{fig:vetobands}
\end{figure*}

In the extreme, the number of outliers produced by a particular instrumental artifact
can be so large as to make systematic follow-up impracticable and pointless.
Table~\ref{tab:outlier_badbands} shows particular frequency bands (about 13\%\ of the original search band) for which no
outlier follow-up is attempted for the initial or later stages because of instrumental artifacts
that lead to such an unmanageable flood of initial or later-stage outliers. The table includes the stage for which the
vetos are applied in order to reduce artifact-induced outlier counts.
Artifacts include loud mechanical resonances, such as higher harmonics of violin modes, and especially
loud hardware injections. The widest bands excluded lie in the regions of test-mass violin modes and their higher harmonics.
Figure~\ref{fig:vetobands} shows these vetoed regions over the full search band, with a magnification of an example
100-Hz band (1900-2000 Hz) that includes a ``forest'' of 4th harmonics of violin modes
from both interferometers.
For the narrow search bands affected by loud hardware injections, only outliers within 1.0 radian
of the injection's sky location are excluded, except for the top 20 (highest SNR),
which are followed up to verify successful injection recovery.

\def\mc#1{\multicolumn{1}{c}{#1}}
\begin{table*}[htb]
\begin{center}
  \begin{tabular}{lrrrr}\hline
    \T\B                                       &    \mc{20-60  Hz}        &  \mc{60-475 Hz}         & \mc{475-1475 Hz}           & \mc{1475-2000 Hz} \\
    \hline\hline
    \T\B Stage 0                                     &  \VlowBandcountzero &  \LowBandcountzero  & \MediumBandcountzero   & \HighBandcountzero \\
    \hline
    \T\B Stage 1                                     &  \VlowBandcountone  &  \LowBandcountone   & \MediumBandcountone    & \HighBandcountone  \\
    \hline
    \T\B Stage 2                                     &  \VlowBandcounttwo  &  \LowBandcounttwo   & \MediumBandcounttwo    & \HighBandcounttwo   \\
    \hline\hline
    \T\B HW injections                               &  \VlowBandcountHWinj  &  \LowBandcountHWinj   & \MediumBandcountHWinj    & \HighBandcountHWinj   \\            
    \T\B Visible artifacts                           &  \VlowBandcountStrainhist  &  \LowBandcountStrainhist   & \MediumBandcountStrainhist    & \HighBandcountStrainhist   \\            
    \T\B Full-O3 follow-up                           &  \VlowBandcountFullO3  &  \LowBandcountFullO3   & \MediumBandcountFullO3    & \HighBandcountFullO3   \\
    \hline\hline
    \T\B Full-O3 survivors                           &  0 & 0 & 0 & 21 \\
    \hline
  \end{tabular}
  \caption{Counts of outliers surviving different stages of the hierarchical
    follow-up for four frequency bands.
    Survivors from stage-2 follow-up are broken down into the categories of hardware injections,
    visible artifacts identified via strain histograms and candidates for loose-coherence follow-up in the
    full-O3 data. Only one cluster of O3a outliers (near 1663.6 Hz) survives the full-O3 fine-grained follow-up, leading to
    a cluster of 21 full-O3 outliers.
    Additional follow-up of these weak candidates, which appear in a visibly disturbed band of L1 data,
    was carried out using the MCMC \pyfstat\ follow-up procedure described in section~\ref{sec:outlier_followup}.
    The \pyfstat\ follow-up confirmed these 21 outliers to be uninteresting.}
\label{tab:outlier_counts}
\end{center}
\end{table*}

\begin{table*}[htb]
  \begin{center}
    \def\mc#1{\multicolumn{1}{c}{#1}}
\begin{tabular}{rrrc@{\hskip0.2in}rrrc@{\hskip0.2in}rrrc}\hline
\mc{$f_{\rm low}^{\rm veto}$} & \mc{$f_{\rm high}^{\rm veto}$} & \mc{$\Delta f^{\rm veto}$} & Stage & \mc{$f_{\rm low}^{\rm veto}$} & \mc{$f_{\rm high}^{\rm veto}$} & \mc{$\Delta f^{\rm veto}$} & Stage & \mc{$f_{\rm low}^{\rm veto}$} & \mc{$f_{\rm high}^{\rm veto}$} & \mc{$\Delta f^{\rm veto}$} & Stage \\ 
\mc{(Hz)} & \mc{(Hz)} & \mc{(Hz)} & applied & \mc{(Hz)} & \mc{(Hz)} & \mc{(Hz)} & applied & \mc{(Hz)} & \mc{(Hz)} & \mc{(Hz)} & applied \\ 
\hline
\hline
25.462 & 25.788 & 0.325 & 2
 & 656.994 & 657.446 & 0.451 & 1
 & 1491.201 & 1501.600 & 10.399 & 0
\\
33.045 & 33.355 & 0.311 & 2
 & 663.264 & 663.716 & 0.453 & 1
 & 1502.660 & 1503.410 & 0.751 & 0
\\
35.539 & 35.876 & 0.336 & 2
 & 666.423 & 666.877 & 0.453 & 1
 & 1503.800 & 1504.700 & 0.901 & 0
\\
38.266 & 38.614 & 0.348 & 2
 & 669.273 & 669.727 & 0.454 & 1
 & 1510.119 & 1513.071 & 2.952 & 0
\\
38.666 & 39.084 & 0.418 & 2
 & 898.520 & 899.510 & 0.990 & 1
 & 1516.948 & 1520.162 & 3.214 & 0
\\
39.256 & 39.664 & 0.408 & 2
 & 899.740 & 900.240 & 0.500 & 1
 & 1527.817 & 1528.773 & 0.956 & 0
\\
40.696 & 41.164 & 0.468 & 2
 & 906.559 & 907.081 & 0.521 & 1
 & 1531.547 & 1532.183 & 0.636 & 0
\\
49.835 & 50.165 & 0.330 & 2
 & 909.839 & 910.361 & 0.522 & 1
 & 1533.507 & 1535.263 & 1.757 & 0
\\
299.200 & 299.980 & 0.780 & 2
 & 918.518 & 919.082 & 0.564 & 1
 & 1537.976 & 1538.614 & 0.638 & 0
\\
301.870 & 302.530 & 0.660 & 2
 & 922.408 & 922.922 & 0.515 & 1
 & 1539.536 & 1540.494 & 0.958 & 0
\\
302.960 & 303.630 & 0.671 & 2
 & 944.586 & 948.745 & 4.159 & 1
 & 1543.856 & 1545.675 & 1.819 & 0
\\
305.819 & 306.581 & 0.761 & 2
 & 946.735 & 947.235 & 0.499 & 1
 & 1549.535 & 1552.165 & 2.630 & 0
\\
307.149 & 307.881 & 0.732 & 2
 & 947.535 & 948.035 & 0.500 & 1
 & 1652.784 & 1653.416 & 0.633 & 2
\\
314.619 & 315.482 & 0.863 & 2
 & 948.235 & 948.735 & 0.500 & 1
 & 1664.514 & 1665.186 & 0.673 & 2
\\
399.800 & 400.190 & 0.390 & 0
 & 958.774 & 959.276 & 0.502 & 1
 & 1666.303 & 1667.037 & 0.733 & 2
\\
409.409 & 411.191 & 1.782 & 0
 & 959.744 & 960.256 & 0.512 & 1
 & 1891.191 & 1891.909 & 0.718 & 1
\\
432.807 & 436.994 & 4.187 & 0
 & 960.744 & 961.256 & 0.512 & 1
 & 1891.961 & 1892.669 & 0.708 & 1
\\
487.001 & 489.599 & 2.598 & 2
 & 962.634 & 963.136 & 0.503 & 1
 & 1922.858 & 1924.262 & 1.405 & 1
\\
492.301 & 517.002 & 24.701 & 2
 & 963.134 & 963.636 & 0.503 & 1
 & 1927.857 & 1929.463 & 1.606 & 1
\\
519.298 & 531.303 & 12.005 & 2
 & 963.834 & 964.336 & 0.503 & 1
 & 1930.237 & 1930.953 & 0.716 & 1
\\
598.740 & 600.220 & 1.480 & 1
 & 964.334 & 966.337 & 2.003 & 1
 & 1930.957 & 1932.793 & 1.836 & 1
\\
604.220 & 604.760 & 0.541 & 1
 & 977.752 & 1051.855 & 74.103 & 0
 & 1935.156 & 1937.844 & 2.687 & 1
\\
606.349 & 606.971 & 0.621 & 1
 & 1082.832 & 1083.368 & 0.537 & 2
 & 1939.256 & 1940.044 & 0.788 & 1
\\
611.889 & 612.811 & 0.922 & 1
 & 1083.432 & 1083.968 & 0.537 & 2
 & 1939.906 & 1941.354 & 1.448 & 1
\\
614.519 & 615.312 & 0.793 & 1
 & 1099.720 & 1100.250 & 0.530 & 2
 & 1944.456 & 1945.194 & 0.739 & 1
\\
629.147 & 630.663 & 1.516 & 1
 & 1152.825 & 1153.375 & 0.551 & 2
 & 1944.705 & 1945.415 & 0.709 & 1
\\
633.087 & 633.543 & 0.457 & 1
 & 1197.970 & 1199.710 & 1.740 & 0
 & 1949.955 & 1963.556 & 13.601 & 1
\\
634.127 & 634.573 & 0.447 & 1
 & 1199.710 & 1200.260 & 0.550 & 0
 & 1965.453 & 1969.847 & 4.394 & 1
\\
638.136 & 638.574 & 0.438 & 1
 & 1208.779 & 1209.921 & 1.142 & 0
 & 1974.253 & 1975.518 & 1.265 & 1
\\
638.436 & 638.884 & 0.448 & 1
 & 1213.179 & 1214.331 & 1.153 & 0
 & 1976.582 & 1977.418 & 0.835 & 1
\\
640.266 & 640.714 & 0.448 & 1
 & 1224.797 & 1225.943 & 1.145 & 0
 & 1977.802 & 1979.238 & 1.436 & 1
\\
642.426 & 642.864 & 0.439 & 1
 & 1229.997 & 1230.943 & 0.946 & 2
 & 1980.342 & 1981.088 & 0.746 & 1
\\
641.976 & 642.424 & 0.448 & 1
 & 1259.424 & 1261.776 & 2.352 & 0
 & 1984.901 & 1986.019 & 1.117 & 1
\\
647.975 & 648.425 & 0.450 & 1
 & 1333.007 & 1333.653 & 0.647 & 2
 & 1988.091 & 1989.519 & 1.428 & 1
\\
648.275 & 648.725 & 0.450 & 1
 & 1458.704 & 1474.297 & 15.593 & 2
 & 1995.815 & 1996.519 & 0.703 & 1
\\
650.935 & 651.385 & 0.450 & 1
 & 1478.752 & 1479.548 & 0.796 & 0
 & 1999.590 & 2000.350 & 0.760 & 1
\\
652.585 & 653.035 & 0.451 & 1
 & 1480.702 & 1483.498 & 2.796 & 0
 & \\
655.954 & 657.406 & 1.451 & 1
 & 1488.961 & 1489.699 & 0.738 & 0
 & \\
\hline
\end{tabular}

    \caption{Frequency bands vetoed from outlier follow-up because of excessive
      outlier counts from instrumental artifacts. The stage at
      which outlier followup is aborted is also shown for each band.
      Altogether, about 257 Hz out of the 1980-Hz original search band was vetoed, or about 13\%.
      Figure~\ref{fig:vetobands} shows these vetoed bands graphically.}
\label{tab:outlier_badbands}
\end{center}
\end{table*}

\begin{figure*}[htbp]
  \ifshowfigs
  \includegraphics[width=6.5in]{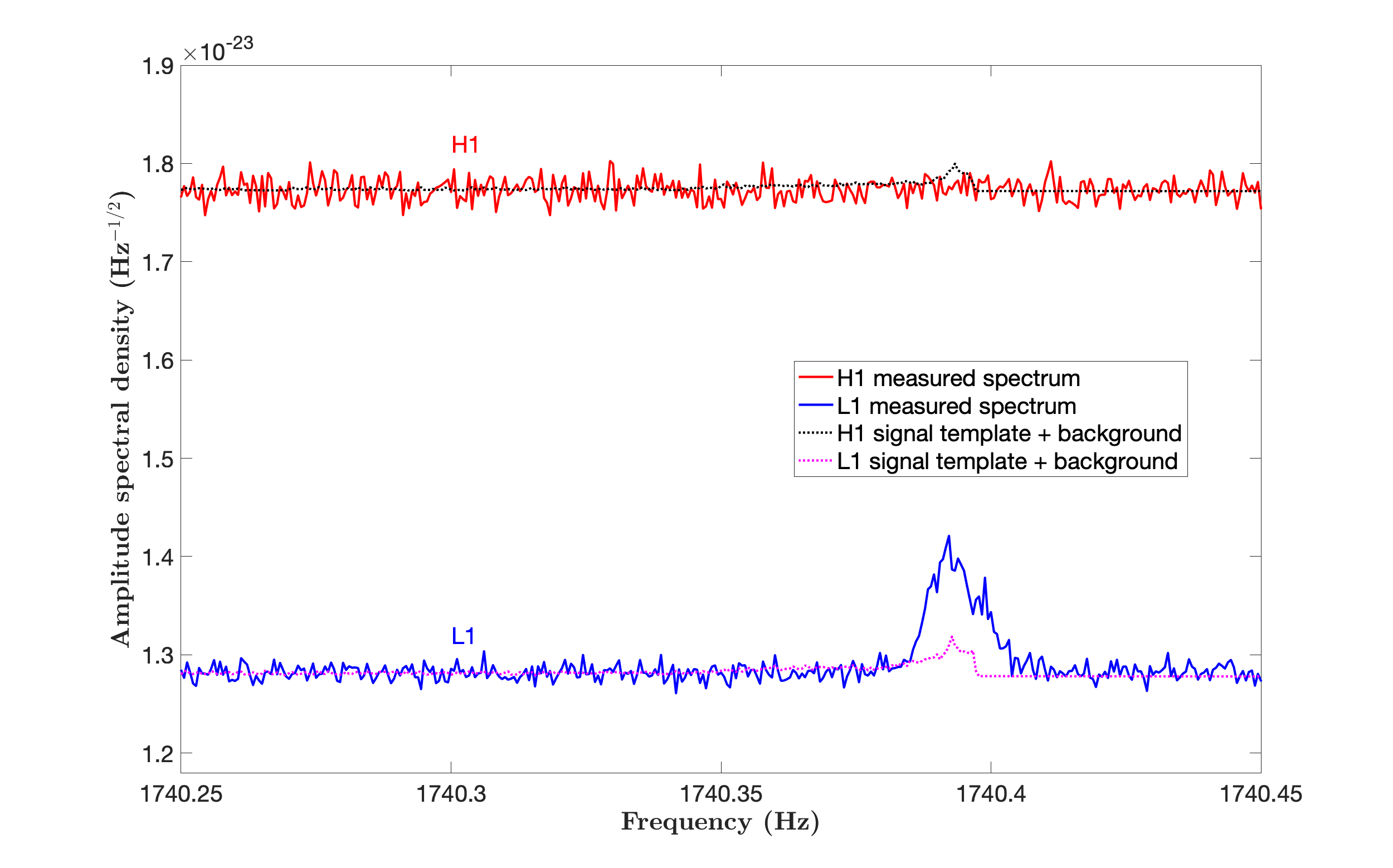}
  \fi
  \caption[Strain Histogram]{
    Example of a ``strain histogram'' graph used in vetoing outliers surviving stage-2 follow-up
    for which instrumental contamination is apparent. The solid curves show the O3a-run-averaged
    H1 (red) and L1 (blue) amplitude spectral densities in a narrow band containing an artifact at 1740.39 Hz.
    The dotted curves show histograms of expected strain excess from H1 (black) and L1 (magenta)
    signal templates added to
    smooth backgrounds interpolated from neighboring frequency bands.
    In this depiction, the strain amplitude of the signal template has been
    magnified to make its structure clear. The large excess power in the L1 data, not seen in the H1 data, despite
    comparable strain sensitivities and comparable sidereal-averaged antenna pattern sensitivities to any point in the sky,
    excludes an astrophysical source for the L1 artifact. The fact that the artifact aligns in frequency
    with the putative signal's template peak in power confirms contamination of the outlier from an
    instrumental source.
    (color online)}
  \label{fig:strainhistogram}
\end{figure*}

Nearly all outliers that survived all stages of the loose-coherence
follow-up correspond to hardware injections (see Table~\ref{tab:hwinjections})
or lie in highly disturbed bands, for which contamination of the putative signal by an instrumental spectral line is
apparent. To identify these contaminations, we construct ``strain histograms'' in which the summed power over the
observation period from a simulation of the nominal
signal candidate is superposed on a background estimate of the noise estimated via interpolation between neighboring
frequency bands.
Except for signal templates with high-magnitude spin-downs, the histograms typically display at least one ``horn''
(narrow peak) from
an epoch during the 6-month O3a period when the orbitally modulated frequency is relatively stationary. We discard
outliers for which the signal template's shape aligns with a spectral artifact known to be instrumental or appearing loudly
in one detector but not the other. Before visual inspection, outliers are clustered in frequency, spin-down and sky location,
with the 10 highest-SNR members receiving scrutiny. For reference, Table~\ref{tab:histogramvetoes} lists the parameters of
the single loudest outlier in each cluster discarded after inspecting strain histograms. Figure~\ref{fig:strainhistogram}
shows an example of a histogram for an outlier contaminated by an L1 spectral line at 1740.39 Hz.

Any remaining survivors (see Table~\ref{tab:followups})
are followed up in the full O3 data set, as described in section~\ref{sec:outlier_followup},
and required to yield an SNR increase of at least 20\%\ relative to the original stage-0 values.
All but one cluster of outliers fail to satisfy this (conservative) requirement 
from the fine-grained full-O3 stage-1 follow-up.

The one cluster with (21) full-O3 outliers passing the requirement occurs near 1663.6 Hz (see Table~\ref{tab:followups}) for
which the maximum SNR increase observed is 29\%.
Examination of the corresponding strain histogram shows evidence of a mild non-Gaussian artifact in L1
data not visible in H1 data, consistent with an instrumental source, but not at a strong enough level to justify visual discarding.

Hence this outlier is subjected
to the MCMC \pyfstat\ follow-up procedure described in section~\ref{sec:outlier_followup}.
The resulting (log) Bayes factor comparing signal to empirical noise expectation 
for the maximum \Fstat\ in the last stage
is found to be log$_{10}$($B_{\rm SN}$) = 1.1. Based on the analysis in~\cite{bib:PyFstatMethod2},
this value is lower by about 50 than that expected for a standard CW signal of the
outlier's putative strain amplitude in the O3a data set.

We conclude that there is no significant evidence for a continuous wave signal from this search.

\begin{table}[htb]
  \begin{center}
    \def\mc#1{\multicolumn{1}{c}{#1}}
\begin{tabular}{rrrr}\hline
\mc{$f$} & \mc{$df/dt$} & \mc{R.A.} & \mc{Dec.} \\ 
\mc{(Hz)} & \mc{(nHz/s)} & \mc{(radians)} & \mc{(radians)} \\ 
\hline
\hline
32.5982 & 0.442 & 4.225 & 0.342
\\
47.5005 & -0.065 & 4.728 & 1.254
\\
53.7245 & -0.473 & 2.142 & -1.107
\\
54.3610 & -1.639 & 4.497 & -1.051
\\
54.8529 & -5.200 & 2.377 & -1.338
\\
64.3629 & -6.569 & 5.266 & -0.906
\\
64.4101 & 0.250 & 0.457 & 1.232
\\
70.0006 & -0.100 & 1.543 & -1.082
\\
78.1135 & 0.638 & 5.688 & 0.078
\\
80.0009 & -0.108 & 4.654 & 1.246
\\
83.3159 & -1.519 & 0.335 & 0.333
\\
85.6917 & -0.546 & 2.454 & -1.212
\\
85.7999 & -8.261 & 2.645 & 0.726
\\
87.9034 & -1.404 & 2.032 & -0.136
\\
89.9920 & 0.569 & 1.051 & -1.485
\\
92.5032 & 1.327 & 4.128 & 0.684
\\
96.0667 & -5.781 & 1.824 & 0.909
\\
99.9780 & -0.585 & 1.363 & 1.151
\\
100.0040 & -0.765 & 1.179 & 1.535
\\
107.1223 & -1.592 & 5.847 & 1.040
\\
368.5864 & -2.796 & 0.133 & -1.017
\\
637.1940 & -2.373 & 0.996 & -0.950
\\
665.9119 & -9.823 & 2.143 & 0.814
\\
1161.6317 & -6.306 & 4.655 & 1.130
\\
1513.4660 & -8.933 & 1.988 & -0.926
\\
1575.8932 & -6.100 & 5.470 & 0.826
\\
1673.5069 & -10.133 & 5.812 & 0.210
\\
1740.4815 & -0.283 & 3.652 & 0.882
\\
1838.6522 & -7.700 & 1.039 & -0.808
\\
\hline
\end{tabular}

    \caption{Parameters of the outliers surviving stage-2 follow-up, but discarded after visual inspection of a strain histogram confirmed instrumental
      contamination. Each outlier listed is the loudest after clustering in frequency, spin-down and sky location. Figure~\ref{fig:strainhistogram} shows
      a strain histogram example.}
\label{tab:histogramvetoes}
\end{center}
\end{table}

\begin{table}[htb]
  \begin{center}
    \def\mc#1{\multicolumn{1}{c}{#1}}
\begin{tabular}{rrrr}\hline
\mc{$f$} & \mc{$df/dt$} & \mc{R.A.} & \mc{Dec.} \\ 
\mc{(Hz)} & \mc{(nHz/s)} & \mc{(radians)} & \mc{(radians)} \\ 
\hline
\hline
329.3400 & -6.669 & 5.231 & 0.531
\\
423.6453 & -8.569 & 5.090 & 1.067
\\
437.0615 & -6.231 & 0.480 & 0.613
\\
444.5386 & -6.131 & 6.136 & 1.160
\\
635.5100 & -7.973 & 3.256 & 0.116
\\
805.0782 & -1.187 & 0.329 & 0.912
\\
1150.9380 & -8.704 & 3.895 & 0.806
\\
1231.9815 & -6.154 & 5.439 & 1.158
\\
1663.6067 & -1.500 & 6.108 & 0.331
\\
\hline
\end{tabular}

    \caption{Parameters of the outliers surviving stage-2 follow-up and followed up in the full-O3 data set.
      Each outlier listed is the loudest after clustering in frequency, spin-down and sky location.}
\label{tab:followups}
\end{center}
\end{table}

\section{\label{sec:conclusions}Conclusions}

We have performed the most sensitive all-sky search to date for continuous gravitational waves
in the range 20-2000\,Hz while probing spin-down magnitudes as high as $10^{-8}$ Hz/s, using the \pf\ search program with loose coherence.
The overall improvements in strain sensitivity come primarily
from the improved noise floors of the Advanced LIGO interferometers over previous LIGO data sets.
Improvements in strain sensitivity over our previous O2 results
range from a factor of 1.2 at $\sim$100 Hz to $\sim$2 at $\sim$2000 Hz, with still larger
improvements at the lowest frequencies for which spectral line artifacts degraded O2 results.

No credible gravitational wave signals are observed, allowing upper limits to be placed on
possible source signal amplitudes.
Fig.~\ref{fig:upper_limits} shows the strain amplitude upper limits obtained.
The lowest upper limits on worst-case (linearly polarized) strain amplitude $h_0$ are  $\sim\sci{1.7}{-25}$ near 200\,Hz.
For a circularly polarized source (most favorable orientation), the smallest
upper limits are  $\sim\sci{6.3}{-26}$.
These upper limits refer to all sky locations and the entire range of frequency derivative values.
For a population-averaged ensemble of sky locations and stellar orientations,
the lowest 95\%\ CL upper limits on the strain amplitude are $\sim\sci{1.4}{-25}$.

At the highest frequencies ($\sim$2000 Hz) we are sensitive to neutron stars with an equatorial
ellipticity $\epsilon$ as small as $\sci{4}{-7}$ and as far away as 6 kpc for
favorable spin orientations (see Fig.~\ref{fig:range}).
For a higher ellipticity $\epsilon$ = $10^{-6}$ and favorable spin orientations,
we are sensitive to neutron stars beyond the galactic center.
The maximum ellipticity that a conventional neutron star can theoretically support
is at least $\sci{1}{-5}$ according to \cite{bib:crust_limit1,bib:crust_limit2}. Our results
are sensitive to such maximally deformed pulsars above $100$\,Hz pulsar rotation
frequency ($200$\,Hz gravitational-wave frequency) within $3$\,kpc.
Outliers from initial stages of each search method are followed up
systematically, but no candidates from any search survived scrutiny.

A recent, similar all-sky search of the LIGO O2 data~\cite{bib:O2Falcon1,bib:O2Falcon2,bib:O2Falcon3},
over the same frequency range (20-2000 Hz),
but severely restricted in spin-down range ($|\dot f|<\sci{3}{-12}$ Hz/s), achieved better strain sensitivity
(improvements ranging from $\sim$30\%\ lower at 500 Hz to $\sim$15\%\ lower at 1500 Hz) by using loose coherence
with an effective coherence time of 12 hours in its initial search stage, at a substantially higher computational cost
than that of this analysis. A similar approach~\cite{bib:deepsearch171-172Hz},
but further restricting the search frequency range to 171--172 Hz,
also achieved a higher sensitivity $h_0 \sim 1.07{\times}10^{-25}$ at 95\%\ confidence.
Figure~\ref{fig:range} illustrates the potential gain in ellipticity and
range sensitivity from searching for high-spin-down magnitudes. A recent deep search of the O2 data using
the distributed-computing project Einstein@Home~\cite{bib:O2EatH} over the frequency band 20-585.15 Hz and
spindown magnitude up to $\sci{2.6}{-9}$ Hz/s achieved similar sensitivity to this search over the common
search parameter space, quoting 90\%\ CL upper limits slightly lower than the 95\%\ CL values presented here.
Figure~\ref{fig:upper_limits} shows the strain upper limits achieved in this O3a search and in the previous O2 searches,
and Fig.~\ref{fig:parameterspace} shows a comparison of their parameter space coverages.

As the LIGO, Virgo and KAGRA gravitational wave detectors improve their strain sensitivities in the coming
decade~\cite{bib:scenariospaper}, searches will probe still smaller neutron star deformations and
explore further out into the galaxy. 

\section{Acknowledgments}
This material is based upon work supported by NSF’s LIGO Laboratory which is a major facility
fully funded by the National Science Foundation.
The authors also gratefully acknowledge the support of
the Science and Technology Facilities Council (STFC) of the
United Kingdom, the Max-Planck-Society (MPS), and the State of
Niedersachsen/Germany for support of the construction of Advanced LIGO 
and construction and operation of the GEO600 detector. 
Additional support for Advanced LIGO was provided by the Australian Research Council.
The authors gratefully acknowledge the Italian Istituto Nazionale di Fisica Nucleare (INFN),  
the French Centre National de la Recherche Scientifique (CNRS) and
the Netherlands Organization for Scientific Research, 
for the construction and operation of the Virgo detector
and the creation and support  of the EGO consortium. 
The authors also gratefully acknowledge research support from these agencies as well as by 
the Council of Scientific and Industrial Research of India, 
the Department of Science and Technology, India,
the Science \& Engineering Research Board (SERB), India,
the Ministry of Human Resource Development, India,
the Spanish Agencia Estatal de Investigaci\'on,
the Vicepresid\`encia i Conselleria d'Innovaci\'o, Recerca i Turisme and the Conselleria d'Educaci\'o i Universitat del Govern de les Illes Balears,
the Conselleria d'Innovaci\'o, Universitats, Ci\`encia i Societat Digital de la Generalitat Valenciana and
the CERCA Programme Generalitat de Catalunya, Spain,
the National Science Centre of Poland and the Foundation for Polish Science (FNP),
the Swiss National Science Foundation (SNSF),
the Russian Foundation for Basic Research, 
the Russian Science Foundation,
the European Commission,
the European Regional Development Funds (ERDF),
the Royal Society, 
the Scottish Funding Council, 
the Scottish Universities Physics Alliance, 
the Hungarian Scientific Research Fund (OTKA),
the French Lyon Institute of Origins (LIO),
the Belgian Fonds de la Recherche Scientifique (FRS-FNRS), 
Actions de Recherche Concertées (ARC) and
Fonds Wetenschappelijk Onderzoek – Vlaanderen (FWO), Belgium,
the Paris \^{I}le-de-France Region, 
the National Research, Development and Innovation Office Hungary (NKFIH), 
the National Research Foundation of Korea,
the Natural Science and Engineering Research Council Canada,
Canadian Foundation for Innovation (CFI),
the Brazilian Ministry of Science, Technology, and Innovations,
the International Center for Theoretical Physics South American Institute for Fundamental Research (ICTP-SAIFR), 
the Research Grants Council of Hong Kong,
the National Natural Science Foundation of China (NSFC),
the Leverhulme Trust, 
the Research Corporation, 
the Ministry of Science and Technology (MOST), Taiwan,
the United States Department of Energy,
and
the Kavli Foundation.
The authors gratefully acknowledge the support of the NSF, STFC, INFN and CNRS for provision of computational resources.
This work was supported by MEXT, JSPS Leading-edge Research Infrastructure Program, JSPS Grant-in-Aid for Specially Promoted Research 26000005, JSPS Grant-in-Aid for Scientific Research on Innovative Areas 2905: JP17H06358, JP17H06361 and JP17H06364, JSPS Core-to-Core Program A. Advanced Research Networks, JSPS Grant-in-Aid for Scientific Research (S) 17H06133, the joint research program of the Institute for Cosmic Ray Research, University of Tokyo, National Research Foundation (NRF) and Computing Infrastructure Project of KISTI-GSDC in Korea, Academia Sinica (AS), AS Grid Center (ASGC) and the Ministry of Science and Technology (MoST) in Taiwan under grants including AS-CDA-105-M06, Advanced Technology Center (ATC) of NAOJ, and Mechanical Engineering Center of KEK.

This document has been assigned LIGO Laboratory document number \texttt{LIGO-P2000334-v9}.

\let\author\myauthor
\let\affiliation\myaffiliation
\let\maketitle\mymaketitle
\title{Authors}
\author{R.~Abbott}
\affiliation{LIGO Laboratory, California Institute of Technology, Pasadena, CA 91125, USA}
\author{T.~D.~Abbott}
\affiliation{Louisiana State University, Baton Rouge, LA 70803, USA}
\author{S.~Abraham}
\affiliation{Inter-University Centre for Astronomy and Astrophysics, Pune 411007, India}
\author{F.~Acernese}
\affiliation{Dipartimento di Farmacia, Universit\`a di Salerno, I-84084 Fisciano, Salerno, Italy  }
\affiliation{INFN, Sezione di Napoli, Complesso Universitario di Monte S.Angelo, I-80126 Napoli, Italy  }
\author{K.~Ackley}
\affiliation{OzGrav, School of Physics \& Astronomy, Monash University, Clayton 3800, Victoria, Australia}
\author{A.~Adams}
\affiliation{Christopher Newport University, Newport News, VA 23606, USA}
\author{C.~Adams}
\affiliation{LIGO Livingston Observatory, Livingston, LA 70754, USA}
\author{R.~X.~Adhikari}
\affiliation{LIGO Laboratory, California Institute of Technology, Pasadena, CA 91125, USA}
\author{V.~B.~Adya}
\affiliation{OzGrav, Australian National University, Canberra, Australian Capital Territory 0200, Australia}
\author{C.~Affeldt}
\affiliation{Max Planck Institute for Gravitational Physics (Albert Einstein Institute), D-30167 Hannover, Germany}
\affiliation{Leibniz Universit\"at Hannover, D-30167 Hannover, Germany}
% ...

\newpage
\maketitle

\end{document}